\documentclass[aps,prb,superscriptaddress,showpacs,twocolumn]{revtex4}
\usepackage[intlimits]{amsmath}
\usepackage{amsfonts}
\usepackage{graphicx}
\usepackage[usenames]{color}
\usepackage{ifpdf}
\ifpdf
\usepackage{epstopdf}
\usepackage[pdftex]{hyperref}
\else
\usepackage[hypertex,ps2pdf,colorlinks,urlcolor=blue,citecolor=blue]{hyperref}
\fi

\begin{document}

\title{XXZ spin-1/2 representation of a finite-$U$ Bose-Hubbard chain at half-integer filling}

\author{Domenico Giuliano}
\affiliation{Dipartimento di Fisica, Universit\`a della Calabria, 
  Arcavacata di Rende I-87036, Cosenza, Italy \& \\
  INFN, Gruppo collegato di Cosenza, 
  Arcavacata di Rende I-87036, Cosenza, Italy}

\author{Davide Rossini}
\affiliation{NEST, Scuola Normale Superiore \& Istituto Nanoscienze-CNR, I-56126 Pisa, Italy}

\author{Pasquale Sodano}
\affiliation{International Institute of Physics, Universidade
  Federal do Rio Grande do Norte, 59012-970, Natal, Brazil \& \\
  INFN, Sezione di Perugia, Via A. Pascoli, I-06123, Perugia, Italy}

\author{Andrea Trombettoni}
\affiliation{CNR-IOM DEMOCRITOS Simulation Center and SISSA, 
  Via Bonomea 265 I-34136 Trieste, Italy \& \\ 
  INFN, Sezione di Trieste, I-34127 Trieste, Italy}

\begin{abstract}
  Using a similarity Hamiltonian renormalization procedure, we determine 
  an effective spin-1/2 representation of the Bose-Hubbard model at 
  half-integer filling and at a finite on-site interaction energy $U$.  
  By means of bosonization, we are able to recast the effective Hamiltonian 
  as that of a spin-1/2 XXZ magnetic chain with pertinently renormalized 
  coupling and anisotropy parameters. We use this mapping to provide analytical 
  estimates of the correlation functions of the Bose-Hubbard model. 
  We then compare such results with those based on DMRG numerical 
  simulations of the Bose-Hubbard model for various values of $U$ and for 
  a number $L$ of lattice sites as low as $L \sim 30$. We find an excellent 
  agreement up to $10\%$ between the output of analytical and numerical 
  computations, even for relatively small values of $U$.
  Our analysis implies that, also at finite $U$, the 1D Bose-Hubbard model 
  with suitably chosen parameters may be seen as a quantum simulator of the XXZ chain. 
\end{abstract}


\maketitle

\section{Introduction}

The study of magnetic systems is one of the most active fields of research 
in condensed matter physics~\cite{blundell01}: the variety of emerging 
ground-states, as well as the rich phase diagram of magnetic lattices, 
makes these systems an optimal testbed to probe 
the competition between various orders and frustration effects~\cite{lacroix11}. 
From this perspective, it would be very useful to be able to engineer 
synthetic physical systems effectively describing magnetic model Hamiltonians, 
with tunable geometry and parameters.

A promising route is provided by cold atomic setups: for instance, itinerant 
magnetism in bulk ultracold Fermi systems with repulsive interactions 
has been experimentally studied~\cite{jo09}, 
while small spin networks have been simulated with ion chains~\cite{kim10}. 
Effective nearest-neighbour spin-spin interactions for atoms 
in neighbour wells of an optical lattice may result 
from super-exchange couplings: 
the corresponding second-order tunneling has been observed in array of double wells~\cite{folling07}.
Furthermore, using fast oscillations of the optical lattice, it is possible to control the sign 
of the nearest-neighbour tunneling~\cite{eckardt05}, which has been recently used 
to simulate classical frustrated magnetism in triangular lattices~\cite{struck11}. 
One may also use two-component gases where the two internal degrees of freedom 
correspond to the simulated (pseudo)spins. Spin interactions can be tuned
by adjusting the external potential~\cite{duan03}.
The recent realization of controllable Bose-Bose mixtures~\cite{thalhammer08} 
paves the way towards the experimental simulation of spin Hamiltonians, in which 
the atomic counterpart of magnetic phases, like antiferromagnetic N\'eel 
and XY ferromagnetic phases (respectively corresponding to the checkerboard 
and the supercounterfluid phases~\cite{kuklov03}) may be detected and studied.  

A key tool in the manipulation of ultracold atomic systems 
is the possibility to superimpose and control optical lattices~\cite{bloch08}. 
The low-energy properties of ultracold bosons in deep optical lattices 
are well captured by the Bose-Hubbard (BH) Hamiltonian:~\cite{jaksch98} 
\begin{equation}
  H_{\rm BH} = \sum_{\langle i,j \rangle} \left[ -t (b^{\dag}_i b_{j} + b^{\dag}_{j} b_i) 
    + V n_i n_{j} \right]  + \frac{U}{2} \sum_{i} n_i (n_i-1) \, .
  \label{HAM_GEN}
\end{equation}
In Eq.~(\ref{HAM_GEN}), $\langle i,j \rangle$ stands for any pair of nearest 
neighbouring sites, while the operators $b^{\dag}_i$ ($b_i$), with 
$[b_i , b_j^\dag ] = \delta_{i,j}$ and $n_i=b^{\dag}_i b_i$, 
create (annihilate) a boson in the site $i$. 
The parameter $t$ denotes the hopping strength, and $U$ ($V$) is the interaction 
energy of two particles at the same site (at two nearest neighbouring sites). 

The use of optical lattices in ultracold atomic systems 
is also central in other proposals to simulate spin Hamiltonians,
such as the quadratic-biquadratic spin model~\cite{garcia04},
or antiferromagnetic spin chains~\cite{sachdev02}.
Following the latter suggestion, by means of a tilted 1D optical lattice,
the Ising chain in a transverse field was experimentally simulated~\cite{simon11}.
The paramagnetic, as well as the antiferromagnetic phase (and the corresponding 
quantum phase transition), were detected by measuring the probability to have an odd 
occupation of sites, while the formation of magnetic domains was observed using in-situ 
site-resolved imaging and noise correlation measurements~\cite{simon11}.

For very large values of $U$, i.e. for $t/U \ll 1$, the BH model 
can be mapped into the Heisenberg XXZ spin-1/2 Hamiltonian:
\begin{equation}
  H_{\rm XXZ} = - J \sum_{\langle i,j \rangle} \left( s_i^x s_{j}^x +
  s_i^y s_{j}^y - \Delta s_i^z s_{j}^z \right)
  \label{xxz_GEN},
\end{equation}
where $\vec{s}_i = \vec{\sigma}_i/2 = \left( s_i^x, s_i^y, s_i^z \right)$ are the $S=1/2$ spin operators,
$\vec{\sigma}_i$ being the Pauli matrices, 
$J$ is the nearest-neighbour coupling, and $\Delta$ is the anisotropy parameter 
($\Delta=\pm 1$ respectively correspond to the antiferromagnetic and the
ferromagnetic isotropic Heisenberg model). 

The use of lattice spin systems for interacting bosons traces back to the classical papers 
by Matsubara and Matsuda in the $50$'s, where the properties of helium II were studied 
assuming that each atom can occupy one of the lattice points~\cite{matsubara56}.
The further assumption that two atoms cannot simultaneously occupy the same lattice site 
(due to the hard-core part of the interparticle interaction between Helium atoms~\cite{aziz79}) 
leads to an effective spin model in a magnetic field~\cite{matsubara56}. 
To qualitatively understand the emergence of a spin representation 
of the one-component BH model one may say that, 
for $U \to \infty$ and {\em if} two states per site give a dominant contribution 
to the energy, an XXZ Hamiltonian is retrieved: this is exactly what happens when the filling $f$, 
defined as the average number of bosons per lattice site, is half-integer. 
Indeed, for $f=\bar{n}+1/2$, with $\bar{n}$ integer, the relevant states 
in the Fock space are given by $| \bar{n} \rangle$ and $| \bar{n}+1 \rangle$ 
(deviations from half-integer fillings would result in a magnetic term 
in the XXZ Hamiltonian). For half-integer $f$, 
at the leading order in $t/U \to 0$ one has $J=2t(f+1/2)$ and $\Delta=V/J$ 
(see the discussion in Sec.~\ref{building}). 

The XXZ model is a paradigmatic spin Hamiltonian which has been the object 
of many investigations and that in 1D is exactly 
solvable by Bethe ansatz~\cite{korepin93,takahashi05}; this provides 
an ideal arena to test different analytical and numerical techniques, 
from bosonization~\cite{gogolin98,giamarchi04} 
to density matrix renormalization group (DMRG)~\cite{schollwock05}. 
The study of (static and dynamical) correlation functions in this model 
is currently an active area of 
research~\cite{kitanine07,pereira07,boos08,essler09,kitanine09,crampe11,klauser11,kozlowski11,sato11,shashi12} 
and exact analytical results for the correlation functions at small distance 
(both at zero and finite temperature) are by now available~\cite{boos08}. 
The asymptotic form of the ground-state correlation functions 
in the thermodynamic limit is power-law with an exponent that has been obtained 
by comparing the result of abelian bosonization with the Bethe ansatz 
solution~\cite{luther75}: for an open chain in the region $-1 \leq\Delta \leq1$, 
the numerical findings for correlation functions obtained with DMRG were compared 
with the results of a low-energy field theory, showing a very good agreement 
and allowing for precise estimates of the amplitudes of the correlation functions~\cite{hikihara98}. 
In turn, the obtained amplitudes were found in agreement with the analytical expressions 
given by Lukyanov and Zamolodchikov~\cite{luk1,luk2}. 
Finally, exact results for the XXZ chain in a special scaling limit 
were used to compute the local correlations 
of a continuous Lieb-Liniger 1D Bose-gas~\cite{pozsgay11}.

In this paper we determine a correspondence between the BH chain 
at half-integer filling for {\em finite} $U$ and a 1D XXZ spin-1/2 model. 
This enables us to provide analytical expressions for the BH correlation 
functions, which we compare with numerical results obtained with DMRG, 
showing that there is a very good agreement both at large and small distances 
and also for $U/J$ as low as $\sim 2$ and for a number of sites $L \geq 30$. 
As a consequence, the numerical determination of the superfluid to charge-density-wave 
and superfluid to Mott-insulator phase transitions 
(respectively corresponding, in the effective XXZ chain, to $\Delta_{\rm eff}=1$ and $\Delta_{\rm eff}=-1$) 
well agrees with the analytical results for the XXZ chain. 
Using our approach, we are able not only to provide analytical expressions for the 1D BH 
correlation functions, but also to show that the BH chain at half-integer filling 
provides a reliable quantum simulator of the XXZ chain. 

In the following we derive an effective spin-1/2 Hamiltonian for the BH chain 
at half-integer filling as a power series of $t/U$. Following Refs.~\onlinecite{glazek93,wegner94}, 
we perform a continuous unitary transformation ${\bf S}$ 
which block-diagonalizes the Hamiltonian in the basis of the eigenvectors 
of $H_{\rm BH}$ with $t=0$ and determine ${\bf S}$ 
perturbatively to the order $(t/U)^2$ (a similar technique
has been used in Ref.~\onlinecite{macdonald88} for the fermionic Hubbard model). 
We finally show that, using bosonization, this Hamiltonian can be recast in the XXZ form 
with pertinent coupling and anisotropy parameters.
We observe that, while to the first order in $t/U$ one finds a XXZ model 
with $J=2t(f+1/2)$ and $\Delta=V/J$, to the next order in $t/U$ 
one gets an effective spin Hamiltonian which is not of the XXZ form, since  
it also contains next-nearest neighbours and $3$-spin terms (this is the 
bosonic counterpart of a similar computation done for the 1D, as well as for the 2D, Fermi-Hubbard 
model~\cite{taka77,chernyshev04,delannoy05}, where $4$-spin terms appear). 
However, in 1D it is possible to proceed further using bosonization:
introducing a Luttinger liquid description of the effective Hamiltonian, we are able 
to incorporate the long-wavelength behaviour of non-XXZ terms 
in the effective coupling and anisotropy parameters, 
$J_{\rm eff}$ and $\Delta_{\rm eff}$,
which are now function of $t$, $V$, $f$ and $U$.

The plan of the paper is the following: after introducing the BH and the XXZ models 
and recalling some useful properties and results (Sec.~\ref{models}), 
we employ the continuous unitary transformation introduced by Glazek and Wilson~\cite{glazek93} 
to approximate the BH chain at half-integer filling with an effective 
spin-1/2 Hamiltonian (Sec.~\ref{building}).
In Sec.~\ref{equivalence} we use bosonization to recast this effective Hamiltonian 
as an XXZ Hamiltonian, with coupling $J_{\rm eff}$ and anisotropy $\Delta_{\rm eff}$, 
while in Sec.~\ref{correco} we establish the correspondence between the 
correlation functions of the BH model and the ones of the XXZ chain. 
We then proceed in comparing the analytical results obtained for the BH 
correlation functions with the numerical findings obtained by DMRG numerical 
simulations (Sec.~\ref{results}), both for the correlation functions and 
the phase transition points. Section~\ref{concl} is devoted to our conclusions, 
while more technical details are contained in the Appendices.

\section{Model Hamiltonians}   \label{models}

Let us start by reviewing the basic properties of the BH and of the spin-1/2 XXZ Hamiltonians, 
in particular focusing on known analytical results about the 
real-space spin correlations in the XXZ chain. 

\subsection{Bose-Hubbard model}  \label{hmodel_0}

The low-energy properties of interacting bosons in a one-dimensional deep optical 
lattice are in general well described by the Bose-Hubbard Hamiltonian~(\ref{HAM_GEN}), which, in 1D 
and with open boundaries, reads: 
\begin{eqnarray}
  H_{\rm BH} &=& -t\sum_{i=1}^{L-1} 
  \left( b^{\dag}_i b_{i+1} + b^{\dag}_{i+1} b_i \right)+
  \frac{U}{2} \sum_{i=1}^L n_i \left(n_i-1 \right) \nonumber \\
&+&
  V \sum_{i=1}^{L-1} n_i n_{i+1}.
  \label{HAM}
\end{eqnarray}
We denote with $N$ the total number of particles in the $L$-site chain, so that the
filling $f$, that is, the average number of particles per site, is given by $f=\frac{N}{L}$. 
For alkali atoms usually $V \ll U$, but with dipolar gases (or polar molecules) $V$ 
could be comparable with $U$: experiments with dipolar gases~\cite{muller11} 
and long-lived ground-state polar molecules~\cite{chotia12} in optical lattices 
have been already performed (see also the review in Ref.~\onlinecite{trefzger11}). 

A large amount of experiments investigated the properties of the BH model: 
the main reason for this interest lies on the fact that this model 
exhibits a quantum phase transition between a superfluid phase 
(for $t/U \gg 1$) and a Mott insulator (for $t/U \ll 1$)~\cite{fisher89}. 
A finite $V$ generally favours charge-density-wave phases: e.g., for half-integer 
filling $f=1/2$, a large $V \gg t, U$ will result in a ground-state 
of the type $| 1,0,1,0,\cdots \rangle$ (where in general 
$| n_1, n_2, n_3, \cdots \rangle$ is an eigenfunction of $H_{\rm BH}$ with $t=0$). 
The ground-state of the BH model has been studied
in the seminal paper in Ref.~\onlinecite{fisher89} using the grand-canonical 
ensemble, where the chemical potential $\mu$ is introduced to enforce 
the constraint on the number of particles. 
The phase diagram in the $U-\mu$ plane shows 
the characteristic lobes: for a pertinently fixed value of $\mu$, the half-integer 
fillings correspond to the ``basis'' of the lobes (i.e. where the lobes touch) and, 
for $V=0$, one has a superfluid for each finite value of $t$, while 
a finite and positive value of $V$ gives rise 
to a charge-density-wave region among the Mott lobes.

The Mott-insulator/superfluid transition was first observed in
3D~\cite{greiner02} and subsequently 
in 1D~\cite{stoferle04} and 2D~\cite{spielman07}. 
The effect of a superimposed external potential (typically a parabolic one) 
has been also considered: the so-called wedding-cake-like 
density has been studied 
both theoretically~\cite{batrouni02,kashurnikov02} and experimentally~\cite{campbell06,folling06}. 
The coherence properties of ultracold bosons in optical 
lattices have been studied, as well, showing that phase coherence 
on short length scales still persists deep in the insulating phase~\cite{gerbier05}.
The BH model in a 1D geometry can be obtained either 
by tightly confining the bosonic cloud in two radial 
directions in presence of a periodic potential in the transverse direction, 
or by creating many (eventually uncoupled) tubes with a 2D optical lattice. 
The properties of strongly correlated phases across the superfluid to Mott-insulator 
phase transition have been analyzed in 1D by means of Bragg spectroscopy~\cite{clement09}. 
The excitation spectrum in the strongly interacting regime has been also studied 
in presence of a tunable disorder, created by a bichromatic optical lattice, 
showing a broadening of the Mott-insulator resonances~\cite{fallani07}.

The finite-$V$ 1D BH model has been studied with a number of analytical and numerical 
techniques: in particular in Ref.~\onlinecite{kuhner00} the phase boundaries of the Mott insulators and 
charge-density-wave phases were determined by DMRG. 
The zero-temperature phase diagram both of the BH model and of a spin-S Heisenberg model 
was constructed and their relation investigated~\cite{amico00}. 
The role of $V$ in inducing supersolid phases in the BH chain was also 
studied~\cite{scarola05,kovrizhin05,batrouni06,mishra09}. 
Bosonization techniques have been applied as well to BH chains, providing a very effective 
way to compute the correlation functions and their decay at large distance~\cite{cazalilla11}).

Finally, we mention that the effect of intersite interactions 
was considered since the 90's in the related quantum phase model, 
describing Josephson junction arrays~\cite{fazio01}: 
this can be obtained from the BH model for large filling per 
site when the number fluctuations are negligible in the kinetic term. 
The chemical potential term in the BH model corresponds to the so-called ``offset charge'' $q$, 
which are external charges present in the superconducting network~\cite{fazio01}: 
the lobes in the quantum phase model are equal, since there is an invariance 
for $q \to q+2e$  ($2e$ being the charge of the Cooper pairs), and 
an half-integer value of the filling $f$ corresponds 
to half-integer values of the offset charges $q/2e$. 
The study of intersite interactions is relevant in Josephson junction arrays since the interaction 
term depends on the capacitance matrix $C_{ij}$, 
which is in general not diagonal, resulting in terms of the form $V_{ij} 
n_i n_j$, where $V_{ij} \propto \left( C_{ij} \right)^{-1}$: 
as a mean-field analysis shows~\cite{grignani00}, for a diagonal 
capacitance matrix one has that at $T=0$ the superconducting phase is obtained 
for each value of the Josephson energy $E_J$ ($\propto t$ in the mapping) and 
that at $q=e$ one has a finite critical temperature for the 
Mott-insulator/superfluid transition for each finite value of $E_J$ (unlike $q=0$, 
where a critical value of $E_J$ is required). Non-diagonal terms of 
the capacitance matrix favour charge density waves~\cite{fazio01}: 
the role of the intersite terms was considered for superconducting chains and 
the corresponding phase diagram investigated~\cite{glark,giuso}, revealing 
that in 1D a (superconducting) repulsive Luttinger liquid phase exists.
The opening of Luttinger liquid phases with tunable parameters
also allows for designing Josephson junction networks supporting emerging 
two-level quantum systems with a high level of quantum coherence
\cite{giuso1,giuso2,cirillo11}.

To conclude this section let us mention that, 
in the rest of the paper, we will mostly
deal with half-integer fillings, $f \equiv \bar{n} +\frac{1}{2}$,
with $\bar{n}=0,1,2,\cdots$. The reason for such a choice is that in this case 
the relevant states for the description of system 
for $U \to \infty$ are just $| \bar{n} \rangle$ and $| \bar{n} + 1 \rangle$. 
Simple arguments, reviewed in Sec.~\ref{XXZmodel_0}, 
then show that, to first order in $t/U$, the BH Hamiltonian is mapped into
an XXZ spin-1/2 Hamiltonian which is integrable in 1D. 
Within the XXZ-model framework, it is also possible to consider small deviations from 
the half-filled regime, which mainly give rise to a uniform magnetic field in the $z$-direction. 
Even though we will not consider large fluctuations in $f$ (of order 1), 
it is possible to take them into account, by keeping, as relevant states for $U \to \infty$, 
$| \bar{n} \rangle$, $| \bar{n} - 1 \rangle$, $| \bar{n} + 1 \rangle$. 
In this case, an effective spin-1 XXZ effective model (in general not integrable) 
is expected~\cite{altman02}. Spin-1 models exhibit a gapped (Haldane) insulator 
phase~\cite{haldane83,shulz1}, which has been investigated in the context 
of the 1D BH model~\cite{dallatorre,amico10,dalmonte11,rossini12}.

\subsection{XXZ chain}   \label{XXZmodel_0}

For a chain with $L$ sites and open boundaries, the Hamiltonian 
of a spin-1/2 XXZ model given in Eq.~(\ref{xxz_GEN}) particularizes to: 
\begin{equation}
  H_{\rm XXZ} = - J \sum_{i=1}^{L-1} \left(s_i^x s_{i+1}^x +
  s_i^y s_{i+1}^y - \Delta s_i^z s_{i+1}^z \right).
  \label{xxz_1}
\end{equation}
The global minus sign in the couplings has been introduced in order to more easily perform 
the comparison with the BH model, and it can be readily gauged away by implementing the canonical mapping 
to the spin-1/2 operators $\tau^a_j$ defined as $\tau_j^{x,y} = (-1)^j s_j^{x,y}$, $\tau_j^z = s_j^z$. 
Therefore the chain is antiferromagnetic (ferromagnetic) for $\Delta$ positive (negative).

Following Ref.~\onlinecite{matsubara56}, one can derive the Hamiltonian in Eq.~(\ref{xxz_1})  
from the BH Hamiltonian~(\ref{HAM}) at half-integer filling $f$ and 
for $U \to \infty$. To do so, let us define $s_j^z \equiv n_j-f$ (so that 
the eigenvalues of $s_j^z$ are $\pm \frac{1}{2}$). 
Since for $t=0$ the energy per particle is (for $L \to \infty$) $\varepsilon = Uf(f-1)/2+Vf^2$,  
it follows that $H_{\rm BH} \stackrel{t \to 0}{\longrightarrow} V s_j^z s^z_{j+1}$, i.e.
\begin{equation}
  J\Delta \equiv V.
  \label{V_0}
\end{equation}
Similarly, for $f \gg 1$, one gets $J\approx 2tf$ 
as one can see by putting $b_i \sim \sqrt{f} e^{i \phi_i}$ 
and mapping the obtained result in the XXZ spin-1/2 language~\cite{glark}: 
for finite values of $f$ one gets (see Sec.~\ref{building}) 
\begin{equation}
  J\equiv 2t \left( f+ \frac{1}{2} \right ).  
  \label{J_0}
\end{equation}
Eqs.~(\ref{V_0}, \ref{J_0}) provide the desired mapping between the BH model 
and the XXZ Hamiltonian to lowest order in  $t/U$. However, as we are going to see 
in Sec.~\ref{correco}, to get a quantitative agreement between the BH and 
the XXZ correlation functions even for $t/U$ relatively small (as low as $0.1$ for $f=1/2$) 
one has to go to the next order in $t/U$: the corresponding Hamiltonian 
is determined in Sec.~\ref{building} and recast in XXZ form via a Luttinger 
representation in Sec.~\ref{equivalence}. 
We remark that, since our result are obtained at half-integer filling, we may omit 
the addition of a magnetic field term of the form $\propto \sum_{i=1}^Ls_i^z$ 
to Eq.~(\ref{xxz_1}). Indeed such a term is proportional to the total spin 
$S_T^z = \sum_{i=1}^Ls_i^z$ in the $z$ direction and, since the system is 
half-filled, only eigenstates of $H_{\rm XXZ}$ with $S_T^z = 0$ 
are physically meaningful - notice that in the following analytical results 
based on the XXZ Hamiltonian~(\ref{xxz_1}) are compared with numerical 
DMRG simulations of the BH chain in the canonical ensemble, where $\sum_i n_i$ 
is conserved and equal to $N$.

The Hamiltonian $H_{\rm XXZ}$ is exactly solvable by means of standard 
Bethe ansatz techniques~\cite{korepin93,takahashi05}: however, explicitly computing 
the real-space spin-spin correlation functions is quite a difficult task.
Exact analytical results for short-range correlators in a range of up to seven lattice sites 
were reported for the isotropic Heisenberg model in Ref.~\onlinecite{sato11}, 
in the thermodynamic limit ($L \to \infty$) and at arbitrary finite temperature, 
and for finite chains of arbitrary length $L$ in the ground-state. 
Results for short-range correlation functions are also available 
for the XXZ chain~\cite{boos08}.
For large distances, using the standard bosonization approach~\cite{gogolin98,giamarchi04} to 
spin-1/2 XXZ model~\cite{bosoniz}, one may find out all the spin-spin correlation functions in terms 
of two-point correlators of pertinent conformal operators~\cite{hikihara98}: in the thermodynamic limit 
one finds the asymptotic forms
\begin{eqnarray}
  \langle \psi_0 | s_i^z s_j^z | \psi_0 \rangle & = & (-1)^{i-j} \frac{A_z}{|i-j|^{1/\eta}}
  -\frac{1}{4 \pi^2 \eta \left( i-j \right)^2},
  \label{therm_1} \\
  \langle \psi_0 | s_i^x s_j^x | \psi_0 \rangle & = & (-1)^{i-j} \frac{A_x}{|i-j|^{\eta}}-\frac{\tilde{A}_x}{|i-j|^{\eta+1/\eta}},
  \label{therm2}
\end{eqnarray}
where $| \psi_0 \rangle$ is the ground-state of $H_{\rm XXZ}$ and we set~\cite{luther75}
\begin{equation}
  \eta =  1 - \frac{1}{ \pi} \arccos{\Delta}.
  \label{eta_gen}
\end{equation}
Analytical expressions for the correlation amplitudes $A_x$, $\tilde{A}_x$ and $A_z$ 
entering Eqs.~(\ref{therm_1}, \ref{therm2}) were presented in Refs.~\onlinecite{luk1,luk2} and 
further discussed in Ref.~\onlinecite{luk3} (see also the discussion in Sec.~V 
of Ref.~\onlinecite{shashi12}): 
\begin{eqnarray}
  A_x &=& \frac{{\cal A}^\eta }{ 8 ( 1 - \eta )^2}  \, e^{- {\cal I}_x} ,                      \label{xxz_6_a}\\
  \tilde{A}_x &=& \frac{{\cal A}^{\eta+1/\eta}}{2 \eta ( 1-\eta ) } \, e^{-\tilde{{\cal I}}_x} ,  \label{xxz_6_b}\\
  A_z  &=&  \frac{2{\cal A}^{1/\eta}}{ \pi^2} \, e^{{\cal I}_z} ,                              \label{xxz_6_c}
\end{eqnarray}
with
\begin{equation} \nonumber
  {\cal I}_x = \int_0^\infty  \frac{dt}{t} 
  \Biggl( \frac{ \sinh ( \eta t )}{ \sinh ( t ) \cosh [ ( 1 - \eta ) t ]} 
  - \eta e^{ - 2 t } \Biggr) ,
\end{equation}
\begin{equation} \nonumber 
  \begin{split}
    \tilde{{\cal I}}_x = \int_0^\infty \frac{dt}{t} \Biggl( & 
    \frac{ \cosh ( 2 \eta t )e^{ - 2 t} - 1}{ 2 \sinh ( \eta t ) \sinh ( t ) \cosh [ ( 1 - \eta ) t ]} \\
    & + \frac{1}{ \sinh ( \eta t )} - \frac{ \eta^2 + 1}{ \eta} e^{ - 2 t } \Biggr) ,
  \end{split}
\end{equation}
\begin{equation} \nonumber
  {\cal I}_z = \int_0^\infty \frac{dt}{t} \Biggl( \frac{ \sinh [ ( 2\eta - 1) t] }
  { \sinh ( \eta t )  \cosh [ ( 1 - \eta ) t ]} - \frac{2 \eta -1}{ \eta} e^{ - 2 t } \Biggr) ,
\end{equation}
and 
\begin{equation}
  {\cal A} = \frac{\Gamma \left( \frac{\eta}{ 2
      (1 - \eta )} \right) }{ 2 \sqrt{\pi} \, \Gamma \left( \frac{1}{ 2 ( 1 - \eta )} \right) } \, ,
\end{equation}
and $\Gamma(x)$ being the Euler's Gamma function.
 
Analytical expressions (in the large-$L$ limit) for the subsequent prefactors 
of the correlation functions are reported in Refs.~\onlinecite{kitanine09,shashi12}.

For chains of finite size $L$ with open boundary conditions, 
one obtains~\cite{hikihara98}:
\begin{widetext}
\begin{eqnarray}
  \langle  \psi_0 | s_i^z s_j^z | \psi_0 \rangle & = & \frac{(-1)^{i-j} a^2}{2f_{\frac{1}{2 \eta}}
    (2 i ) f_{\frac{1}{2 \eta}} (2 j ) } \: \left( \frac{  f_{\frac{1}{\eta}} (i+ j )}{
    f_{\frac{1}{\eta}} (i- j ) } - \frac{  f_\frac{1}{\eta} (i- j) }{ f_\frac{1}{\eta} (i+ j ) } \right) 
  - \frac{1 }{ 4 \pi^2 \eta} \left( \frac{1}{f_2 ( i-j) } +  \frac{1}{f_2 ( i+j) } \right) \nonumber
  \\
  & & - \frac{a }{ 2 \pi  \eta} \Biggl\{ \frac{ (-1)^i}{ f_{\frac{1}{2 \eta}}
    (2 i ) } [ g (i-j)+ g (i+j) ] - \frac{ (-1)^j}{ f_{\frac{1}{2 \eta}} (2 j ) } [ g (i-j)- g (i+j) ] \Biggr\}
  \label{xxz_2}
\end{eqnarray}
and
\begin{eqnarray}
 \langle  \psi_0 | s_i^x s_j^x | \psi_0 \rangle & = & 
 \frac{ f_\frac{\eta}{2} ( 2 i ) f_\frac{\eta}{2} ( 2 j ) }{
    f_\eta ( i - j ) f_\eta ( i +j ) } \: \Biggl\{ (-1)^{i-j} \frac{c^2}{2} 
   - \frac{ b^2}{ 4 f_\frac{1}{2 \eta}
    (2 i ) f_\frac{1}{2 \eta}
    (2 j ) } \left[ \frac{f_\frac{1}{\eta} ( i+j)}{  f_{\frac{1}{\eta}} ( i-j)}
    + \frac{f_{\frac{1}{\eta}} ( i-j)}{f_{\frac{1}{\eta}} ( i+j)} \right] \nonumber \\
  &&- \frac{bc}{2} \: {\rm sgn} ( i - j ) \left[ \frac{ (-1)^i}{  f_{\frac{1}{2 \eta}}
      (2 j ) } - \frac{ (-1)^j}{  f_{\frac{1}{2 \eta}}
      (2 i ) }  \right] \Biggr\}, 
  \label{xxz_4}
\end{eqnarray}
\end{widetext}
where ${\rm sgn} ( x )$ is the sign function and 
\begin{eqnarray}
  f_\alpha ( x ) &=& \left[ \frac{ 2 ( L + 1 )}{ \pi} \sin \left( \frac{ \pi | x | }{ 2 (
      L + 1 )} \right) \right]^\alpha, \label{xxz_3_a} \\
  g ( x ) &=&   \frac{ \pi}{2 ( L + 1) } \cot \left( \frac{ \pi  x  }{ 2 (
    L + 1 )} \right) ,
  \label{xxz_3_b}
\end{eqnarray}
with 
\begin{equation}
  \frac{c^2}{2} \equiv A_x, \;\; \frac{b^2}{4} \equiv \tilde{A}_x, \;\;
  \frac{a^2}{2} \equiv A_z
  \label{xxz_5}
\end{equation}
(here and in the following all the distances are in units of the lattice constant). 

The agreement between exact numerical calculations of the XXZ correlation
functions and analytical expressions in~(\ref{xxz_2}, \ref{xxz_4}) is very good, 
and it becomes excellent with $L \sim 100$ for $-0.8 \lesssim \Delta \lesssim 0.8$~\cite{hikihara98}. 
Thus one may readily assume that Eqs.~(\ref{xxz_2}, \ref{xxz_4}) provide quite an accurate 
analytical expression for the spin-spin correlation functions in the
XXZ model~\cite{nota_0}. As a consequence, constructing a rigorous mapping between 
the BH and the XXZ spin-1/2 Hamiltonian and expressing correlation
functions of one model in terms of the ones of the other model gives an efficient 
and straightforward way to provide accurate analytic expressions for real-space 
correlation functions in the BH model at half-integer filling. 

We finally observe that the only system-dependent parameter determining the spin-spin correlation functions 
is the coefficient $\eta$: thus, in tracing out the mapping between the two models, 
this is the key quantity to be calculated as a function of the BH parameters. 
In particular, one may distinguish between the regions in parameter space with
$\eta > 1/2$ and $\eta < 1/2$: while the former one corresponds to an antiferromagnetic spin chain, 
the latter one (which may be realized for pertinently chosen values of the parameters of $H_{\rm BH}$, 
as we shall show below) corresponds to a ferromagnetic chain.

\section{Effective Spin-1/2 Hamiltonian for the Bose-Hubbard model at half-integer filling}   \label{building}

As reviewed in the previous section, for $U \to \infty$, the BH Hamiltonian 
maps onto the XXZ model in Eq.~(\ref{xxz_1}), with the parameters $J , \Delta$ 
given in Eqs.~(\ref{V_0}, \ref{J_0}). 
This may be seen as a first-order term in an expansion (in powers of $t/U$) 
aimed at computing the effective Hamiltonian: in this section we compute 
this effective Hamiltonian to the next order. As we shall show in the following, 
this is enough to fit quite well the numerical data for the correlation 
functions of the BH model using the analytical results obtained 
for the correlators of the XXZ chain.

To approach the large-$U$ limit one may either proceed by performing a strong 
coupling expansion to the second or higher-order of perturbation theory, 
or by deriving effective Hamiltonians using alternative techniques, 
based on canonical transformations or continuous unitary transformations~\cite{mila_chap}.
At integer filling, for instance, it is possible to evaluate the energy of the 
Mott insulator and of the superfluid state in higher-order perturbation theory and determine 
the phase diagram in the $U-\mu$ plane~\cite{freericks96}. 
Since we are rather interested to the BH at half-integer filling, i.e., in
the region of the phase diagram where the lobes touch and the superfluid phase 
persists also at very small $U$ (with $V=0$), we found it convenient to use 
an approach based on continuous unitary transformations~\cite{glazek93,wegner94}. 
We follow the notation and the method presented in the paper 
by Glazek and Wilson (GW)~\cite{glazek93}: systematically using the GW 
renormalization procedure, we work out an effective description of the dynamics 
of the BH model, restricted to the low-energy subspace determined by the constraint 
on the total number of particles and by the large-$U$ assumption. 
As a result, the low-energy subspace is spanned by states with either $\bar{n}$ 
or $\bar{n} + 1$ particles per site, with the total number of particles 
being fixed to $N$. Thus, the space of physically relevant states
at each site is in one-to-one correspondence with the Hilbert space of
states of a quantum spin-1/2 degree of freedom; we shall see that, at 
half-integer filling, even for finite $U$ the BH model may be replaced 
by an effective spin-1/2 Hamiltonian, with pertinently determined parameters. 
The method amounts to an iterative block-diagonalization of the BH Hamiltonian 
on the space of eigenfunctions of $H_{\rm BH}$ with $t=0$.

To illustrate the procedure, we start from the explicit construction 
of the ``low-energy'' 
Hilbert space of physically relevant states, in the large-$U$ limit. 
Neglecting excitations with energy $\sim U$ amounts to  
truncating the Hilbert space to a subspace ${\cal F}$, defined as
\begin{equation}
  {\cal F} = {\rm Span} \{ | \bar{n} + \mu_1 , \ldots , \bar{n} + \mu_L \rangle \},
  \label{spinsemi1}
\end{equation}
with $\mu_i$ taking the values  $\mu_i= 0 , 1$ and $\sum_{i = 1}^L \mu_i = \frac{L}{2}$. 
In Eq.~(\ref{spinsemi1}) $|n_1 , \cdots , n_L \rangle$ labels the state in the Hilbert space
with $n_i$ particles on site $i$. To implement the GW approach, one splits the 
Hamiltonian~(\ref{HAM}) as $H_{\rm BH} = H_0 + H_I$, with 
\begin{eqnarray}
  H_0 & = & \frac{U}{2} \sum_{i} n_i ( n_i - 1 ) + V \sum_{ i } n_i n_{i+1}
  \label{model_3_a} \\
  H_I & = & -t \sum_i \left( b^{\dag}_i b_{i+1} + b^{\dag}_{i+1} b_i \right).
  \label{model_3}
\end{eqnarray}
From Eqs.~(\ref{model_3_a}, \ref{model_3}) one sees that $H_0$ is diagonal with respect to the partition of
the Hilbert space into ${\cal F}$ plus its orthogonal complement, since
\begin{equation}
  H_0 | n_1 , \ldots , n_L \rangle = E_0 [ n_1 , \ldots , n_L ] | n_1 , \ldots , n_L \rangle
  \label{model_5}
\end{equation}
with $E_0 [ n_1 , \ldots , n_L ] = (U/2)  \sum_{ i } n_i ( n_i - 1 ) + V \sum_i n_i n_{i+1}$, 
while $H_I$ exhibits off-diagonal (with respect to the partition
of the Hilbert space) matrix elements which are ${\cal O} (t\bar{n})$. 
In order to block-diagonalize $H_{\rm BH}$, one needs to perform a {\em similarity}
transformation~\cite{glazek93} 
\begin{equation}
  H_{\rm BH}  \to  {\cal \tilde{H}}_{\rm BH} = {\bf S}^\dag H_{\rm BH} {\bf S},
  \label{model_6}
\end{equation}
with ${\bf S}$ unitary. 
Upon setting ${\bf S} = {\bf I} + {\bf T}$, the unitarity of ${\bf S}$ implies the optical theorem
\begin{equation}
  {\bf T} +  {\bf T}^\dag +  {\bf T}^\dag {\bf T} = 0.
  \label{simi_1}
\end{equation}
Setting ${\bf T} \equiv {\bf h} + {\bf a}$, with  
\begin{equation}
  {\bf h} = \frac{1}{2} \left( {\bf T} +  {\bf T}^\dag \right), \,\,\,
  {\bf a} = \frac{1}{2} \left( {\bf T} -  {\bf T}^\dag \right)
  \label{simi_2}
\end{equation}
one finds that Eq.~(\ref{simi_1}) yields
\begin{equation}
  {\bf h} = \frac{1}{2} ({\bf a}^2 - {\bf h}^2 ).
  \label{simi_3}
\end{equation}
Eq.~(\ref{simi_3}) shows that ${\bf h}$ is always ``higher order'' than ${\bf a}$. 
Following Ref.~\onlinecite{glazek93}, it is most convenient to define 
the new interaction Hamiltonian ${\cal \tilde{H}}_I$ as
\begin{equation}
  {\cal \tilde{H}}_I = {\cal \tilde{H}}_{\rm BH} - H_0
  \label{simi_4}
\end{equation}
so that the new ``free'' Hamiltonian is the same as the old one ($H_0$). 

To further proceed and determine ${\bf S}$, one has to require that the matrix elements of ${\cal \tilde{H}}_I$ 
between states with energy difference  $\gtrsim U$ are equal to zero, amounting to state that 
${\cal \tilde{H}}_I$ is block-diagonal with respect to the partition of the Hilbert space into
${\cal F}$ plus its orthogonal complement, i.e. 
\begin{equation}
  {\cal P} {\cal \tilde{H}}_I {\cal P} + \left( {\bf I} - {\cal P} \right)
  {\cal \tilde{H}}_I \left( {\bf I} - {\cal P} \right) = {\cal \tilde{H}}_I, 
  \label{simi_5}
\end{equation}
where ${\cal P}$ is the projector onto ${\cal F} $ and ${\bf I} - {\cal P}$
the projector onto its complementary subspace~\cite{nota_1}. 
One sees that Eq.~(\ref{simi_5}) implies that
\begin{equation}
  {\cal P} {\cal \tilde{H}}_I \left( {\bf I} - {\cal P} \right) 
  = \left( {\bf I} - {\cal P} \right) {\cal \tilde{H}}_I {\cal P} = 0.
  \label{simi_6}
\end{equation}
Using Eqs.~(\ref{model_6}, \ref{simi_2}, \ref{simi_4}), one may write ${\cal \tilde{H}}_I$ as
\begin{equation}
  {\cal \tilde{H}}_I = \left( {\bf I} + {\bf h} - {\bf a} \right) \left( H_0 + H_I \right) 
  \left( {\bf I} + {\bf h} + {\bf a} \right) - H_0
  \label{simi_add}
\end{equation}
and Eq.~(\ref{simi_6}) then becomes
\begin{equation}
\begin{split}
  {\cal P} \{ H_I + \{ H_0 , {\bf h} \} & + [ H_0 , {\bf a} ] + {\bf T}^\dag H_I +H_I {\bf T} \\
  &+ {\bf T}^\dag H_I {\bf T}   \}  \left( {\bf I} - {\cal P} \right) = 0.
  \label{simi_7}
\end{split}
\end{equation}
Eq.~(\ref{simi_7}), together with the  identity 
\begin{equation}
  {\bf a} =  {\cal P}  {\bf a} \left( {\bf I} - {\cal P} \right) + 
  \left( {\bf I} - {\cal P} \right) {\bf a}  {\cal P} 
  \label{simi_ter}
\end{equation}
and with Eq.~(\ref{simi_3}), is all what one needs in principle to fully determine ${\bf a}$ and ${\bf h}$ 
(and, therefore, the operator ${\bf T}$). 

However, except for some simple cases~\cite{wegner94}, an explicit solution for ${\bf T}$ cannot be exhibited. 
For this reason we proceed by writing the solution for ${\bf T}$ iteratively, in a series in $H_I$: 
in particular, we use Eq.~(\ref{simi_7}) to determine ${\bf a}$ to first order (${\bf a_1}$) in $H_I$. 
We provide the details in Appendix~\ref{appgw} and the result for ${\bf a_1}$ in Eq.~(\ref{simi_12}). 
Using Eq.~(\ref{simi_12}) and setting ${\bf T} \approx {\bf a_1}$, we find that Eq.~(\ref{model_6}) reads   
\begin{equation}
  {\bf S}^\dag H_{\rm BH} {\bf S} = H_{\rm BH} + [ H_0 , {\bf a_1}  ] + [ H_I , {\bf a_1} ].
  \label{ex_1}
\end{equation}
The GW procedure may be readily iterated to determine, in principle, ${\bf T}$ to any
desired order in $H_I$. However, since keeping only second-order contributions
in $H_I$ provides already quite an excellent estimate for the real-space correlation functions
of operators in the BH model (as explicitly shown by the numerical calculations 
we report in Sec.~\ref{results}), setting ${\bf T} \approx {\bf a_1}$ 
already provides quite a good approximation to the exact ${\bf T}$. 

Since the approach we are implementing is perturbative in $H_I$, 
one should enforce Eq.~(\ref{simi_7}), as well as Eq.~(\ref{simi_ter}), to
each order in $H_I$; moreover, since ${\cal P}  [ H_0 , {\bf a_1}  ] {\cal P}  = 0$, 
one may neglect the term $[ H_0 , {\bf a_1} ]$ in Eq.~(\ref{ex_1}) and approximate 
the effective Hamiltonian acting within ${\cal F} $ as
\begin{equation}
  H_{\rm eff} = {\cal P} \left\{ H_{\rm BH} + [ H_I , {\bf a_1}  ] \right\} {\cal P} 
  \equiv H_{\rm XXZ}^{(0)} + H^{(1)}.
  \label{ex_final}
\end{equation}
The first term in the right hand side of Eq.~(\ref{ex_final}) yields a spin-1/2 Hamiltonian 
which is actually the spin-1/2 XXZ chain introduced in Sec.~\ref{XXZmodel_0} 
and having the anisotropy and the coupling given by Eqs.~(\ref{V_0}, \ref{J_0}): 
\begin{equation}
 H_{\rm XXZ}^{(0)} \equiv {\cal P} H_{\rm BH} {\cal P}  = 
- J \sum_{i } \hspace{-1mm} \left( \hspace{-1mm} s_i^x s_{i+1}^x + 
  s_i^y s_{i+1}^y -  \frac{V}{J} s_i^z s_{i+1}^z \hspace{-1mm} \right) \hspace{-1mm} 
  \label{spinsemi2}
\end{equation}
with $J = 2 t \left( f + \frac{1}{2} \right)$ (constant terms have been omitted). 
The effective spin-1/2 operators are defined as
\begin{eqnarray}
  s_i^x &=& \frac{1}{2 \sqrt{f + \frac{1}{2}}} {\cal P}  \left( b_i + b_i^\dag \right) {\cal P}, \qquad \nonumber \\
  s_i^y &=& \frac{i}{2 \sqrt{f + \frac{1}{2}}} {\cal P}  \left(- b_i + b_i^\dag \right) {\cal P}, \qquad \nonumber \\
  s_i^z &=& {\cal P}  \left( b_i^\dag b_i - f \right) {\cal P};
  \label{spinsemi3}
\end{eqnarray}
the boson number eigenstates at site $i$ correspond to the eigenstates of $s_i^z$ according
to  $|\bar{n} \rangle_i  \leftrightarrow | \downarrow \rangle_i$,
and $|\bar{n} + 1 \rangle_i \leftrightarrow | \uparrow \rangle_i$.
Therefore, the result in Eq.~(\ref{spinsemi2}) corresponds to
the ``naive'' large-$U$ limit for the BH model at half-integer filling discussed 
in Sec.~\ref{XXZmodel_0}, in which off-diagonal matrix elements of relevant operators (including
the Hamiltonian itself) are set to zero from the very beginning. 

Corrections to $H_{\rm XXZ}^{(0)}$ arising from virtual transitions involving states outside of ${\cal F} $
may be properly accounted for within GW procedure, allowing to 
get the effective 
spin-1/2 Hamiltonian to the next order in $t/U$. Summing over all virtual transitions 
outside of ${\cal F} $ induced by $H_I$, one finds
\begin{eqnarray}
  &&H^{(1)}   \equiv {\cal P}   [ H_I , {\bf a_1} ]{\cal P}  = 
  - t^2 \sum_{ j , \ell } {\cal P} \left( b_j^\dag b_{j + 1} + b_{j + 1}^\dag b_j \right) \times \nonumber \\
  &&  \times (  {\bf I} - {\cal P}  )  
  \left( H_{\rm BH} \right)^{-1}
  \left( {\bf I} - {\cal P} \right) \left( b_\ell^\dag b_{\ell + 1} + b_{\ell + 1}^\dag b_\ell \right) {\cal P}.
\label{spinsemi5}
\end{eqnarray}
In particular, when computing $H^{(1)}$, one has to consider intermediate states 
with either one of the $\mu_j$ in Eq.~(\ref{spinsemi1}) being equal to $2$, or to $-1$
(all these states have energy $\sim U$, with respect to states in the
subspace ${\cal F}$), or states with one of the $\mu_j$ equal to $2$ ($-1$),
and the other equal to $-1$ ($2$) (all these states have energy $\sim 2 U$, 
with respect to states in the subspace ${\cal F}$). Thus, one eventually finds out 
that $H^{(1)}$ can be written as the sum of two terms: 
$H^{(1)} = H_{\rm diag}^{(1)} + H_{\rm offd}^{(1)}$, with 
$H_{\rm diag}^{(1)}$ being the part of $H^{(1)}$ having 1- and 2-nearest-neighbour spin terms, 
while $H_{\rm offd}^{(1)}$ contains 2-next-nearest-neighbour and 3-spin terms. 
Omitting constant terms, their expression are given by:
\begin{widetext}
  \begin{eqnarray}
    H_{\rm diag}^{(1)}  & = & - \frac{4 ( \bar{n} + 1 ) t^2}{U} \sum_{ i } s_i^z 
    - \frac{t^2}{U} \left( 3 \bar{n}^2 + 6 \bar{n} + 4 \right) \sum_{ i } s_i^z s_{i + 1}^z
    \label{map1.2} \\
    H_{\rm offd}^{(1)}  & = & - \frac{t^2 ( \bar{n} + 1)^2}{U} 
    \sum_{i} \left( s_{i+1}^+ s_{i-1}^- + s_{i+1}^- s_{i-1}^+ \right)
    - \frac{2 t^2 ( \bar{n} + 1)}{U} \sum_{i} \left( s_{i+1}^+ s_{i-1}^- + s_{i+1}^- s_{i-1}^+ \right) s_i^z.
    \label{map1.3}
  \end{eqnarray}
\end{widetext}
As we shall see in the next section, using a Luttinger liquid representation, 
$H^{(1)}$ may be recast in the XXZ form with coupling and anisotropy 
coefficients depending on $U$.

\section{Effective XXZ parameters via a Luttinger liquid representation}   \label{equivalence}

The effective spin Hamiltonian in Eq.~(\ref{ex_final}) is not in the XXZ form: 
in this section we show how the contribution coming from $H^{(1)}$ may be accounted 
for by a pertinent redefinition of the parameters of the spin-1/2 XXZ-Hamiltonian $H_{\rm XXZ}^{(0)}$. 

The first contribution to $H_{\rm diag}^{(1)}$ in the right-hand side of Eq.~(\ref{map1.2}) 
describes an effective magnetic field in the $z$ direction~\cite{nota_2}, 
while the second term simply shifts the value of the XXZ anisotropy. 
At variance, the term $H_{\rm offd}^{(1)}$ in Eq.~(\ref{map1.3}) 
contains $3$-spin, as well as non-nearest neighbour, couplings. 
To show how these terms can be accounted for via a redefinition of $H_{\rm XXZ}^{(0)}$, 
it is most convenient to introduce the Jordan-Wigner (JW) 
fermions $a_j , a_j^\dag$~\cite{gogolin98}, 
in terms of which one gets 
\begin{equation}
  H_{\rm XXZ}^{(0)}  =  - 2 J \sum_k \cos ( k ) a_k^\dagger a_k 
  + J \Delta \sum_j : \hspace{-1mm} a_j^\dag a_j \hspace{-1mm} : \: 
  : \hspace{-1mm} a_{j+1}^\dag a_{j+1} \hspace{-1mm} : \, ,
  \label{adiz.1}
\end{equation}
where $a_k$ are JW fermionic operators in momentum space and $: \: :$ denotes
normal ordering with respect to the fermionic ground-state. In terms of 
JW fermions, one  writes $H_{\rm offd}^{(1)}$ as a sum of a bilinear ($H_2$), 
plus a quartic ($H_4$) term, that is $$H_{\rm offd}^{(1)} \equiv H_2 + H_4,$$ with
\begin{eqnarray}
  H_2 \hspace{-0.5mm} & = & \hspace{-0.5mm} \frac{t^2 ( \bar{n} + 1 )}{U}
  \sum_{ i } \left( a_{i-1}^\dag a_{i+1} + a_{i + 1}^\dag a_{i-1} \right) 
  \label{map1.4_2} \\
  H_4 \hspace{-0.5mm} & = & \hspace{-0.5mm} \frac{2 t^2 ( \bar{n} +1 )^2}{U} \hspace{-1mm}
  \sum_{ i } \hspace{-0.5mm} : \hspace{-1mm} a_i^\dag a_i  \hspace{-1mm} : \hspace{-0.5mm}
  \left( a_{i-1}^\dag    a_{i+1} + a_{i+1}^\dag    a_{i-1} \right) .
  \label{map1.4}
\end{eqnarray}
Since $H_2$ is bilinear in the JW fermions, it merely modifies the
single-fermion dispersion relation, yielding 
the quadratic Hamiltonian in the JW fermions reading
\begin{eqnarray}
  H_{\rm XXZ}^{(0)} + H_2 & = & \sum_k \bigg\{ - 2 J \cos{k} + \frac{t^2 ( \bar{n} + 1 )}{U} \cos{(2 k)} \nonumber\\
  & & - B \bigg\} a_k^\dag a_k \, ,
  \label{map1.6}
\end{eqnarray}
with $B = 4 ( \bar{n} + 1 ) t^2 / U$. Setting $\epsilon ( k )= - 2 J \cos{k}
+ \frac{t^2 ( \bar{n} + 1 )}{U} \cos{(2 k)} -B $, one finds that the Fermi
points, defined by $\epsilon ( k_F ) = 0$, are given by
\begin{equation}
  \cos{k_F} = \frac{U ( \bar{n} + 1)}{2 J } - \sqrt{\left( \frac{U  (\bar{n} + 1)}{2 J} \right)^2 + \bar{n} +2}.
  \label{map1.7}
\end{equation}
Upon linearizing the dispersion relation around $\pm k_F$ and setting $k = k_F + p$, one gets
\begin{equation}
  \epsilon ( \pm k_F + p ) \approx \pm J \sin{k_F} \left[ 1 - \frac{2J}{U  (\bar{n} + 1)} \cos{k_F} \right]  \, p.
  \label{map1.8}
\end{equation}
From Eq.~(\ref{map1.8}) one sees that, since $\cos{k_F}\neq 0$, $H_2$ implies 
a nonzero effective magnetic field $B_{\rm eff}$~\cite{nota_2}, as well as 
a redefinition of the Fermi velocity $v_F$. This yields a redefined coupling 
given by $B_{\rm eff} / J_{\rm eff} = -\cos{k_F}$. Since 
$$
B_{\rm eff}=-J \cos{k_F} \left( 1 - \frac{2J}{U} \cos{k_F} \right),
$$
one obtains
\begin{equation}
  J_{\rm eff} =  J \left( 1 - \frac{2J}{U} \cos{k_F} \right).
  \label{map1.10}
\end{equation}
The quartic term $H_4$ can be dealt with by noticing that, 
in the low-energy, long-wavelength limit, one can write
\begin{eqnarray}
 && a_{j-1}^\dag a_{j+1} + a_{j+1}^\dag a_{j-1}  \longrightarrow - \biggl\{ \rho_R ( x_j ) + \rho_L ( x_j ) \nonumber \\
&&  - (-1)^j [ \psi_R^\dag ( x_j ) \psi_L ( x_j )  +  \psi_L^\dag (x_j ) \psi_R ( x_j ) ] \biggr\} \,,
  \label{map1.12}
\end{eqnarray}
where the chiral fermion fields $\psi_R ( x_j)$, $\psi_L ( x_j )$ are
defined from the long-wavelength expansion of $a_j$ as 
\begin{equation}
  a_j \approx e^{ i k_F x_j} \psi_R ( x_j ) + e^{ - i k_F x_j} \psi_L ( x_j ),
  \label{map_ultragat}
\end{equation}
with $x_j = aj$, and the chiral fermion  densities  given by 
$\rho_R ( x_j ) = \psi_R^\dag ( x_j ) \psi_R ( x_j )$ and $\rho_L ( x_j ) =  \psi_L^\dag ( x_j ) \psi_L ( x_j )$. 
As a result, $H_4$ may be written as
\begin{eqnarray}
  H_4 & = &  - \frac{ 4 t^2 ( \bar{n} + 1)^2}{U} \: \int_0^L \: 
  d x \:   \Big\{ ( \rho_R ( x ) )^2 \nonumber \\
  & & + ( \rho_L ( x  ) )^2  + 4 \rho_R ( x  ) \rho_L ( x  ) \Big\}. 
  \label{map1.13}
\end{eqnarray}
Comparing Eq.~(\ref{map1.13}) to Eq.~(\ref{adiz.1}), one sees that 
$H_4$ takes the same form as the term $J \sum_{j} s_j^z s_{j+1}^z$
in the spin-1/2 XXZ Hamiltonian in Eq.~(\ref{xxz_1}). 

\begin{figure}[t]
  \includegraphics[width=\columnwidth]{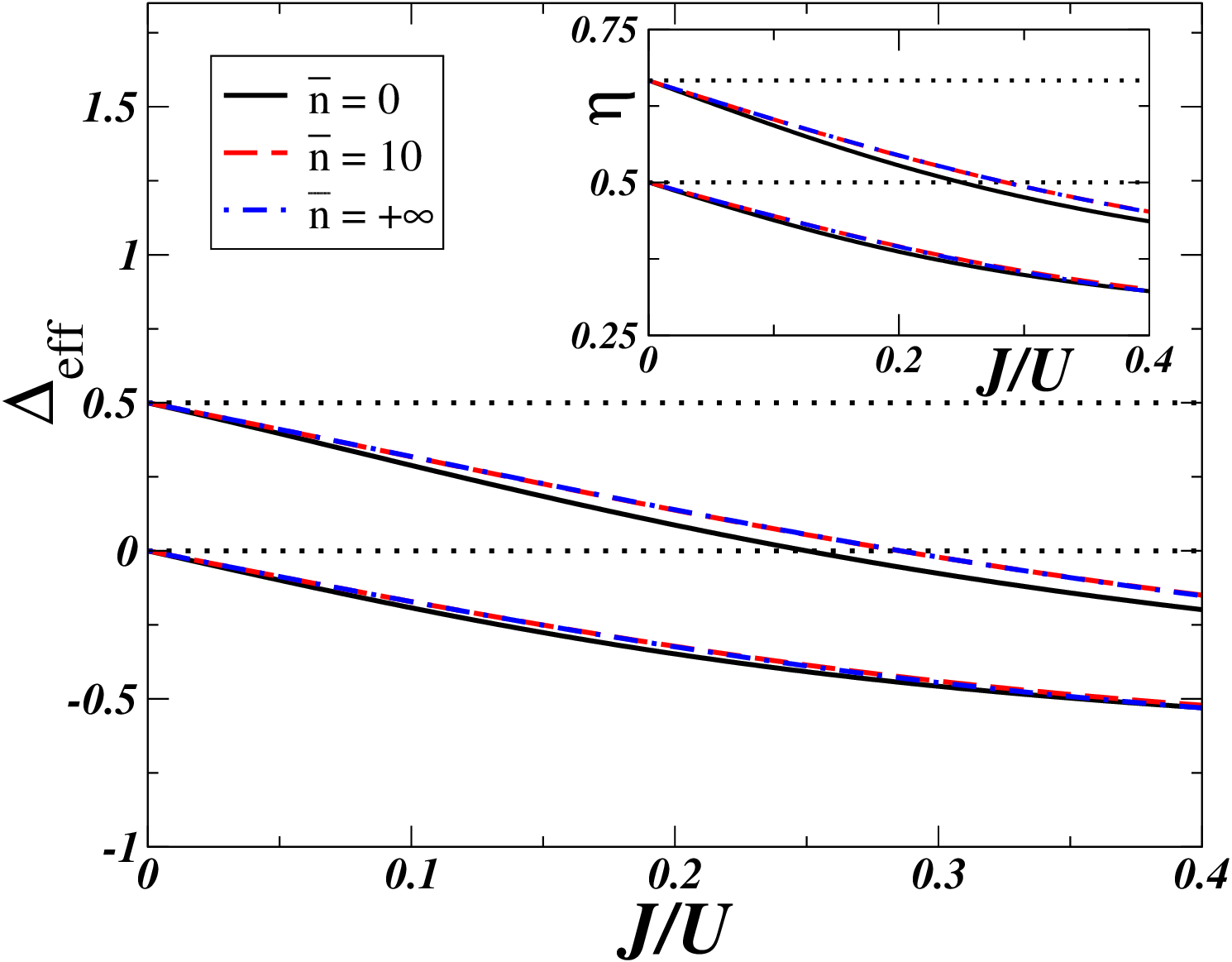}
  \caption{$\Delta_{\rm eff}$ {\it vs.} $J/U$ for different values
    of $V/J$ and $\bar{n}$. The top (bottom) dotted line corresponds to 
    the value of $\Delta_{\rm eff}$ for $U/J \to \infty$ and $V / J = 0.5$ ($V / J = 0$). 
    The other lines are for $V/J=0.5$ (top) and $V/J=0$ (bottom), with $\bar{n}=0$ (solid black lines), 
    $\bar{n}=10$ (dashed red lines) and $\bar{n} \to \infty$ (dot-dashed blue blue lines).
    Inset: same as in the main panel, but for $\eta$ {\it vs.} $J/U$.}
  \label{FIGURE1}
\end{figure}

Collecting together all the above results allows to write an effective XXZ Hamiltonian, 
describing the BH model to the order $(t/U)^2$, as:
\begin{equation}
  H_{\rm XXZ}^{\rm eff} = - J_{\rm eff} 
  \sum_{ j } \left( s_j^x s_{j+1}^x + s_j^y s_{j+1}^y - \Delta_{\rm eff}s_j^z s_{j+1}^z \right),
  \label{map1.18}
\end{equation}
with $J_{\rm eff}$ defined in Eq.~(\ref{map1.10}) and 
\begin{equation}
  \Delta_{\rm eff} = \frac{\bar{\Delta}}{1- \frac{2J}{U} \cos (k_F)} 
  \label{map1.17}
\end{equation}
with
\begin{equation}
  \bar{\Delta} = \frac{V}{J} - \frac{ t^2 ( 3 \bar{n}^2 + 6 \bar{n} + 4)}{JU} - \frac{ 4 t^2 ( \bar{n} + 1)^2}{JU}.
  \label{map1.15}
\end{equation}

Since $J_{\rm eff}$ acts just as an effective over-all scale of $H_{\rm XXZ}^{\rm eff}$, 
then $\Delta_{\rm eff}$ is the only parameter determining the behavior of spin-spin correlations 
in the XXZ model. Substituting Eq.~(\ref{map1.17}) in Eq.~(\ref{eta_gen}) 
one gets
\begin{equation}
  \eta  = 1 - \frac{1}{\pi} \arccos{\Delta_{\rm eff}},
  \label{eta_BH}
\end{equation}
which provides an explicit formula for the effective Luttinger parameter 
for the BH model at half-integer filling. In Fig.~\ref{FIGURE1} we plot 
both $\Delta_{\rm eff}$ and $\eta$ versus $J/U$, for different values of $V/J$ 
and $\bar{n}$. One sees that $\bar{n}=10$ and $\bar{n} \to \infty$ are almost 
indistinguishable, and that the limit of the quantum phase model for Josephson 
junction arrays ($\bar{n} \gg 1$) at offset charge $q=e$ 
is practically reached at $\bar{n} \sim 10$. Furthermore, one sees 
that the dependence of $\eta$ upon $\bar{n}$ is rather small. 

From Fig.~\ref{FIGURE1} one also sees that $\Delta_{\rm eff}$ may be tuned 
by varying the ratio $J/U$: in particular $\Delta_{\rm eff}$ can be different 
from $0$ even if $V=0$ (as it is typical for alkali atoms). 
Fig.~\ref{FIGURE1} also suggests the possibility of describing the
whole phase diagram of the XXZ spin-1/2 chain using BH model for a 
single species of bosons with pertinently chosen parameters, 
see also Sec.~\ref{results}~\cite{nota_3}.

Finally we notice that, since the sign of $\Delta_{\rm eff} $ may be changed 
by a pertinent choice of $J/U$ and $V$, the Luttinger liquid effectively
describing the XXZ-Hamiltonian may be repulsive or attractive. 
As noticed in the context of 1D Josephson junction arrays~\cite{glark,giuso}, 
the transition between the repulsive and the attractive side 
may be monitored by inserting a weak link (i.e., a nonmagnetic impurity~\cite{bosoniz}): 
it would be then interesting to analyze the effects of a weak link 
introduced in a bosonic system described by the BH Hamiltonian.

\section{Correlation functions}   \label{correco}

The mapping between $H_{\rm BH}$ and $H_{\rm XXZ}^{\rm eff}$ derived in 
Sec.~\ref{equivalence} enables to select the ground-states on 
which to compute the pertinent vacuum expectation values.
Indeed if $ | \Phi_0 \rangle$ is the ground-state of the 
BH Hamiltonian given in Eq.~(\ref{HAM}), 
and $ | \Psi_0 \rangle \equiv {\bf S^\dag} | \Phi_0 \rangle$ is 
the ground-state of $H_{\rm eff} = {\bf S}^\dag H_{\rm BH} {\bf S}$,  
the GW approach requires 
\begin{eqnarray}
  \langle \Phi_0 | {\cal O}_{\rm BH} \left[ \left\{ b , b^\dag \right\} \right]  
  \Phi_0 \rangle & = & 
  \langle \Psi_0 | {\bf S}^\dag {\cal O}_{\rm BH} 
  \left[ \left\{ b , b^\dag \right\} \right]  {\bf S} | \Psi_0 \rangle \nonumber \\
  & \equiv &  \langle \Psi_0 |  {\cal O}_{\rm XXZ} 
\left[ \left\{ s^a \right\} \right] |  \Psi_0 \rangle,
\label{corco.1}
\end{eqnarray}
where ${\cal O}_{\rm BH} \left[ \left\{ b , b^\dag \right\} \right]$ 
(${\cal O}_{\rm XXZ} \left[ \left\{ s^a \right\} \right]$) denotes a generic BH (XXZ) operator.
Of course, Eq.~(\ref{corco.1}) is exact only if ${\bf S}$ is the exact solution of the GW 
equation~(\ref{simi_ter}): by computing it perturbatively at a given order, 
one recovers the correspondence between ground-state expectation values 
of BH and spin-1/2 operators at the chosen order. 

In the rest of the paper, we will be interested in correlation functions of the following BH operators:
\begin{eqnarray}
  {\cal M}_{i,j}^z    & \equiv & \left( n_i - f \right) \left( n_j - f \right), \label{corko.a} \\
  {\cal M}_{i,j}^\perp & \equiv & b_i^\dag b_j. 
\label{corko.b}
\end{eqnarray}
Using the results of Appendix~\ref{appop} one has 
${\bf S}^\dag {\cal M}_{i,j}^z {\bf S} = 
{\cal M}_{i,j}^z \left[ 1 + {\cal O} \left( \frac{t^2 \bar{n}^2}{U^2} \right) \right]$, so that
\begin{equation}
\langle \Phi_0 | \left( n_i - f \right) \left( n_j - f \right) | \Phi_0 \rangle =
\langle \Psi_0 | s_i^z s_j^z |  \Psi_0 \rangle 
+  {\cal O} \left( \frac{t^2 \bar{n}^2}{U^2} \right). 
\label{corco.12}
\end{equation}
More generally, if the operator ${\cal O}_{\rm BH}$ satisfies 
$\left( {\bf I} - {\cal P} \right) {\cal O}_{\rm BH} {\cal P} = 
{\cal P} {\cal O}_{\rm BH} \left( {\bf I} - {\cal P} \right) = 0$, then 
$\langle \Phi_0 | {\cal O}_{\rm BH}  | \Phi_0 \rangle  \approx   
\langle \Psi_0 | {\cal O}_{\rm XXZ} | \Psi_0 \rangle$, with ${\cal O}_{\rm XXZ}$
obtained from ${\cal O}_{\rm BH} $ by substituting $b_i$, $b_i^\dag$ and 
$n_i - f$ respectively with $s_i^-$, $s_i^+$ and $s_i^z$. 
At variance, for ${\cal M}_{i,j}^\perp$ one obtains a more involved expression 
(see Appendix~\ref{appop} for details):
\begin{widetext}
  \begin{eqnarray}
    \label{corco.17}
    \langle \Phi_0 | b_i^\dag b_j | \Phi_0 \rangle & \approx &
    (\bar{n} + 1 )  \langle \Psi_0 |  s_i^- s_{j}^+ | \Psi_0 \rangle +
    \frac{t ( \bar{n} + 2)(\bar{n} + 1 ) }{2 U} \langle \Psi_0 |  [ s_{i + 1}^- 
      + s_{i- 1}^-  ] s_{j}^+  +
    \left[ s_{j + 1}^+ + s_{j - 1}^+ \right]  s_{i}^- | \Psi_0 \rangle \\ 
    \nonumber & &
    + \frac{t  \bar{n} (\bar{n} + 1 ) }{2 U} \langle \Psi_0 | s_{i }^-  [ s_{j -1}^+ 
      +  s_{j +1}^+  ]+ 
    s_{j  }^+ \left[ s_{i -1}^- +  s_{i +1}^- \right]  | \Psi_0 \rangle +
    \delta_{ | i - j | , 1} \: \langle \Psi_0 |  \left( \frac{1}{2} - s_{i + 1}^z \right) 
    \left( \frac{1}{2} + s_{i }^z  \right) | \Psi_0 \rangle,
  \end{eqnarray}
\end{widetext}
where again we neglected contributions arising to ${\cal O} \left( \frac{t^2 \bar{n}^2}{U^2} \right)$.

\section{Results}   \label{results}

In this section we compare the numerical results obtained by means of DMRG 
for the correlation functions and the phase diagram of the BH model 
with the analytical predictions for the correlators from the effective Hamiltonian 
$H_{\rm XXZ}^{\rm eff}$ given by Eq.~(\ref{map1.18}).

\subsection{Correlation functions}   \label{corr_sec}

Let us focus on the BH correlation functions. 
Since DMRG simulations are performed on a finite number of sites $L$ 
and for open boundary conditions, we may use Eqs.~(\ref{xxz_2}, \ref{xxz_4}) 
yielding the $zz$ and $xy$ correlation functions of the XXZ model. 
We evaluate the values of the non-universal constants $a$, $b$, $c$ defined 
in Eq.~(\ref{xxz_5}) both numerically and analytically, by using 
the expressions presented in Refs.~\onlinecite{luk1,luk2} and reported in 
Sec.~\ref{XXZmodel_0}. 
As confirmed in Ref.~\onlinecite{hikihara98}, the values of $a$, $b$, $c$ obtained in the two ways 
are in excellent agreement. 
We show that the analytical expressions for the XXZ correlations are 
well confirmed by the numerical BH correlations also for small $L$ (e.g., for 
$L=30$) and for $J/U$ relatively large (as large as $\sim 0.5$). It should be 
stressed that, at variance, the agreement is not very good 
by setting $\Delta_{\rm eff}= V / J $, i.e. by using the Hamiltonian $H_{\rm XXZ}^{(0)}$ 
obtained for $U \to \infty$ neglecting contributions arising from the GW procedure. 

\begin{figure}[b]
  \includegraphics[width=\columnwidth]{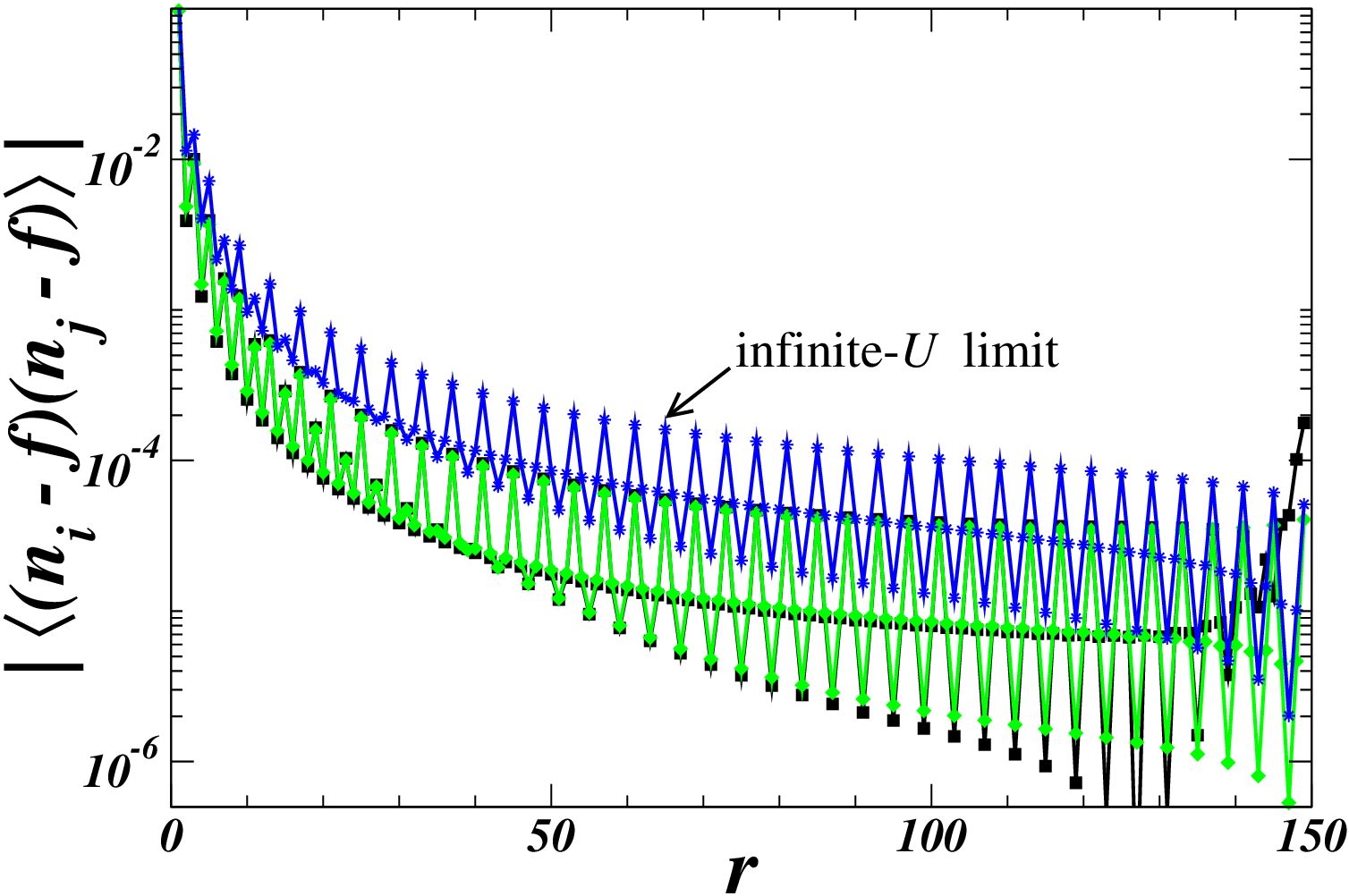}
  \caption{Density-density correlations 
    $\left| \langle \left(n_i-f \right)\left(n_j-f \right)\rangle \right|$ 
    {\it vs.} $r=\left| i-j \right|$ for $U=10t$, $V=0.5t$ and $f=0.5$, 
    with number of sites $L=150$. Black squares: numerical BH results; 
    green diamonds: XXZ result with $a$, $b$, $c$ analytically determined; 
    blue stars: $U \to \infty$ XXZ result (indicated by the label 
    ``infinite-$U$ limit'' -- see text fur further details). Lines are guide for eye. 
    On the scale of the figure, results obtained for the XXZ model with $a$, $b$, $c$ 
    numerically determined (not shown here) are indistinguishable from the ones 
    obtained with the corresponding analytical values. 
    Notice also the excellent agreement between numerical BH findings 
    and analytical XXZ results.} 
  \label{FIGURE2}
\end{figure}

\begin{figure}[t]
  \includegraphics[width=\columnwidth]{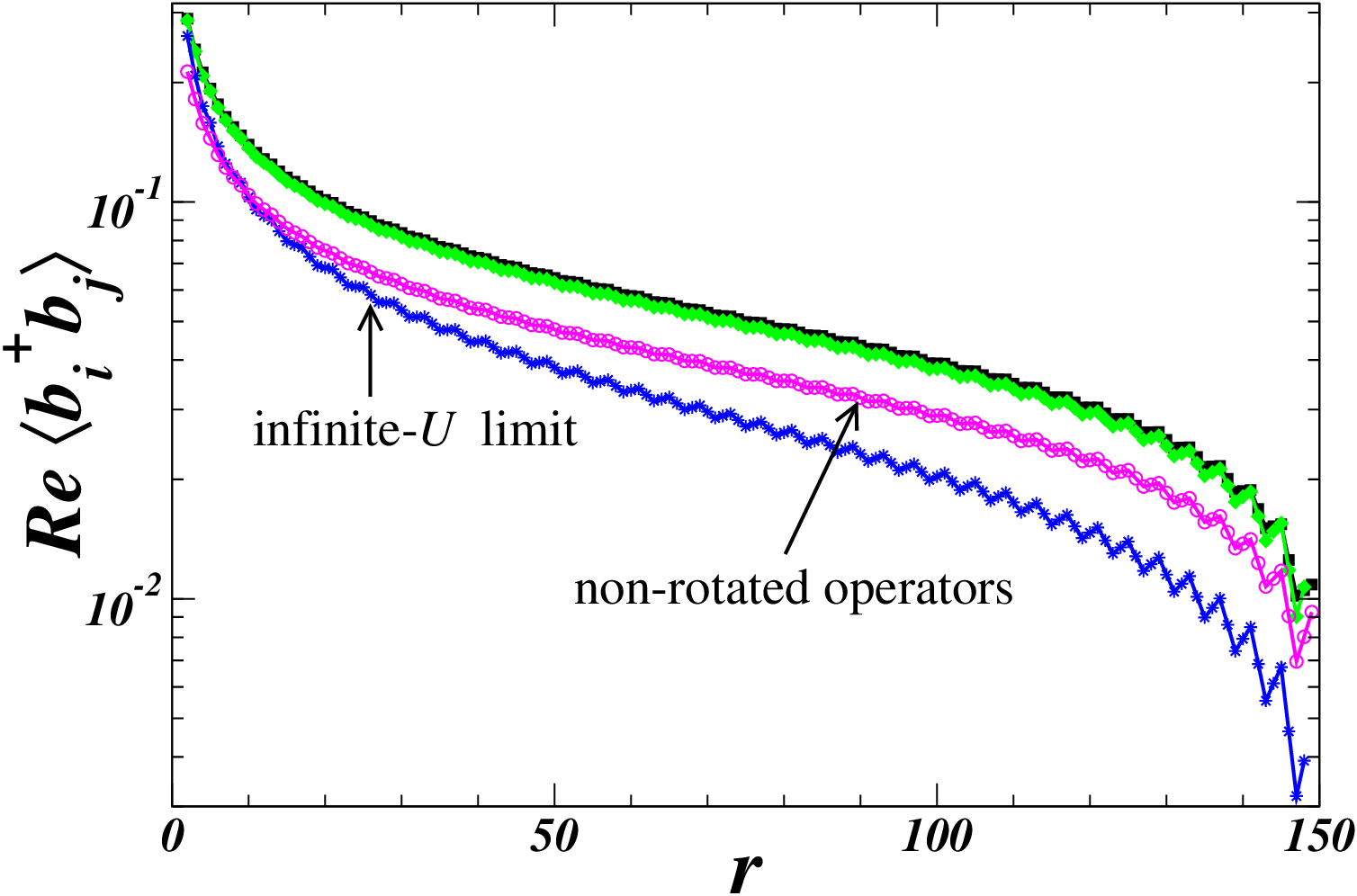}
  \caption{Real part of $\langle b_i^\dag b_j \rangle$
    {\it vs.} $r=\left| i-j \right|$ for $U=10t$, $V=0.5t$, $f=0.5$, $L=150$. 
    Magenta circles denote XXZ results with ``non-rotated operators''. 
    The notation for the other symbols is the same as in Fig.~\ref{FIGURE2}.}
  \label{FIGURE3}
\end{figure}

\begin{figure}[t]
  \includegraphics[width=\columnwidth]{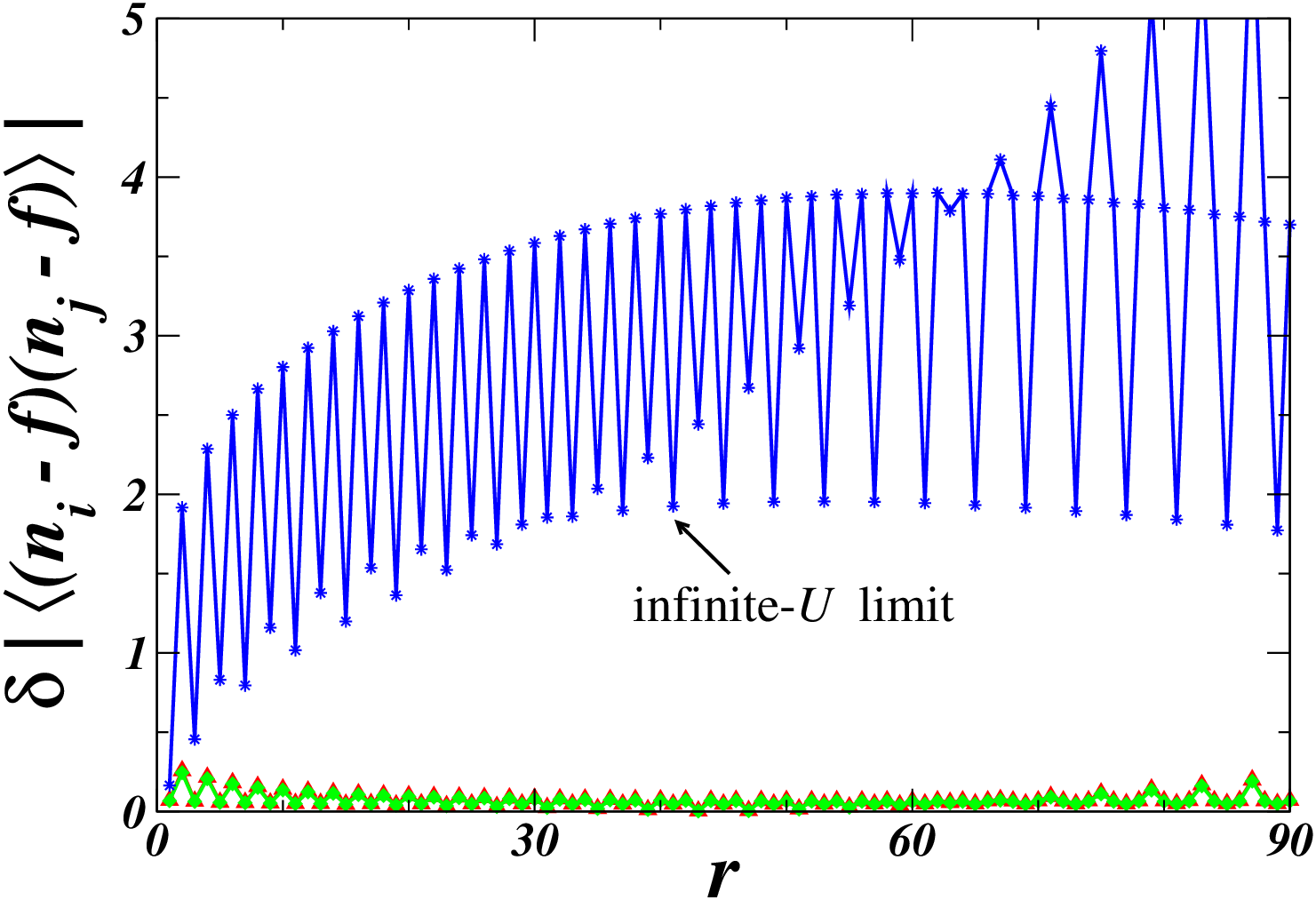}
  \caption{Relative error of the correlation function 
    $\left| \langle \left(n_i-f \right)\left(n_j-f \right)\rangle \right|$ 
    {\it vs.} $r=\left| i-j \right|$ for the same parameters (and the same conventions 
    for symbols and lines) of Fig.~\ref{FIGURE2}. We also plot the results 
    obtained from the XXZ model with $a$, $b$, $c$ numerically determined as red triangles: 
    numerical and analytical estimates for finite $U$ practically coincide.  
    The average value, with $r_{\rm max}=3L/5$, is $0.06 \pm 0.04$ for the finite-$U$ XXZ model  
    and $2.9 \pm 1.2$ for the infinite-$U$ XXZ model.} 
  \label{FIGURE4}
\end{figure}

\begin{figure}[t]
  \includegraphics[width=\columnwidth]{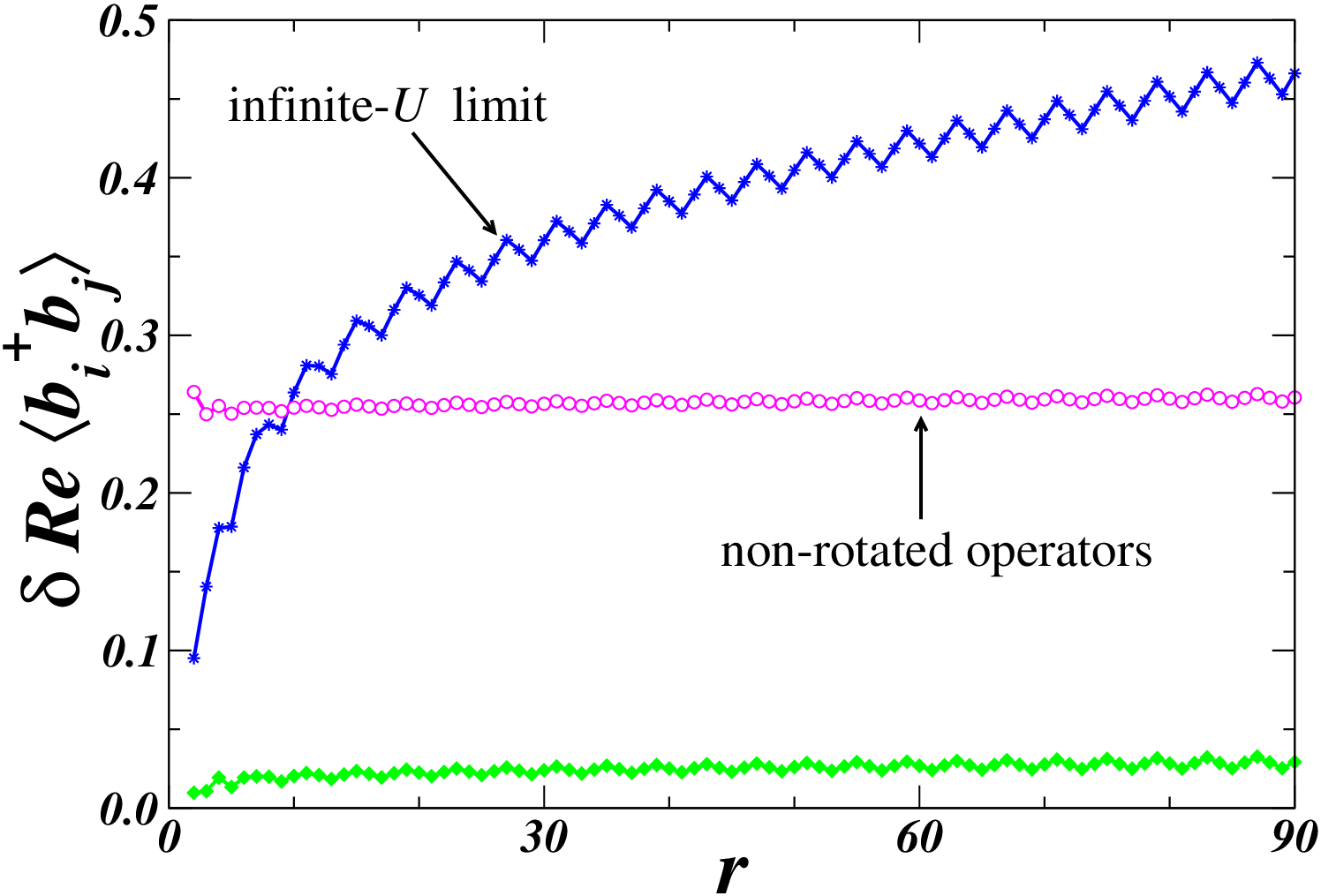}
  \caption{Relative error of the real part of $\langle b_i^\dag b_j \rangle$ 
    {\it vs.} $r=\left| i-j \right|$ for the same parameters (and the same conventions for symbols and lines) 
    of Fig.~\ref{FIGURE3}. The average values, with $r_{\rm max}=3L/5$, are: 
    $0.025\pm0.004$ (green diamonds - analytical XXZ results), 
    $0.38\pm0.08$ (blue stars - XXZ result in the infinite-$U$ limit), 
    $0.26\pm0.03$ (magenta circles - finite-$U$ result with non-rotated operators).} 
  \label{FIGURE5}
\end{figure}

\begin{figure}[t]
  \includegraphics[width=\columnwidth]{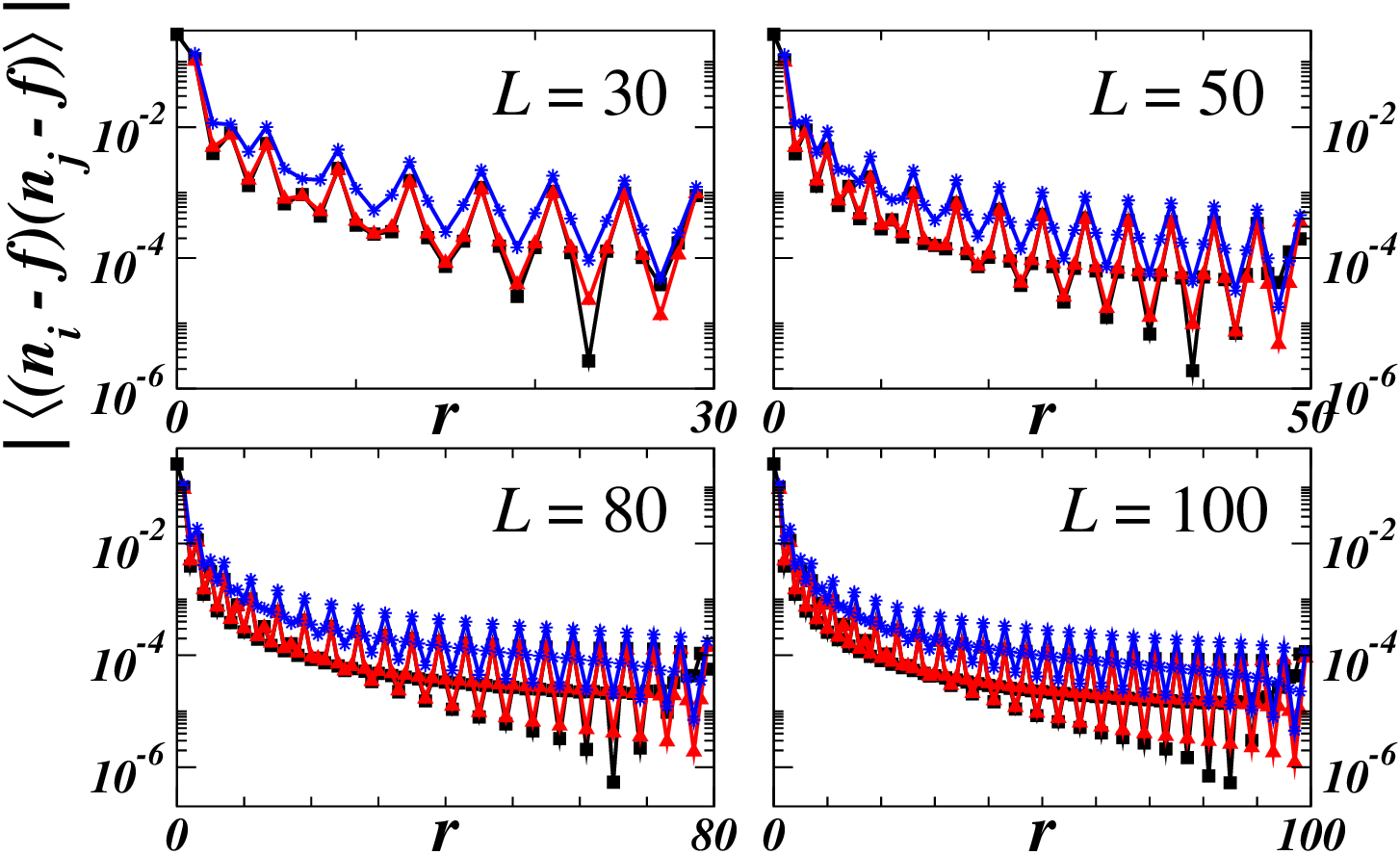}
  \caption{In each panel we plot $\left| \langle \left(n_i-f \right)\left(n_j-f \right)\rangle \right|$ 
    {\it vs.} $r=\left| i-j \right|$ for different sizes: $L=30,50,80,100$ 
    ($U=10t$, $V=0.5t$, $f=0.5$). Black squares denote the BH results, 
    red triangles the XXZ results with $a$, $b$, $c$ numerically determined, 
    and blue stars the infinite-$U$ XXZ results (we do not report the XXZ results 
    with $a$, $b$, $c$ analytically determined, since they practically coincide 
    with the red triangles).} 
  \label{FIGURE6}
\end{figure}

\begin{figure}[t]
  \includegraphics[width=\columnwidth]{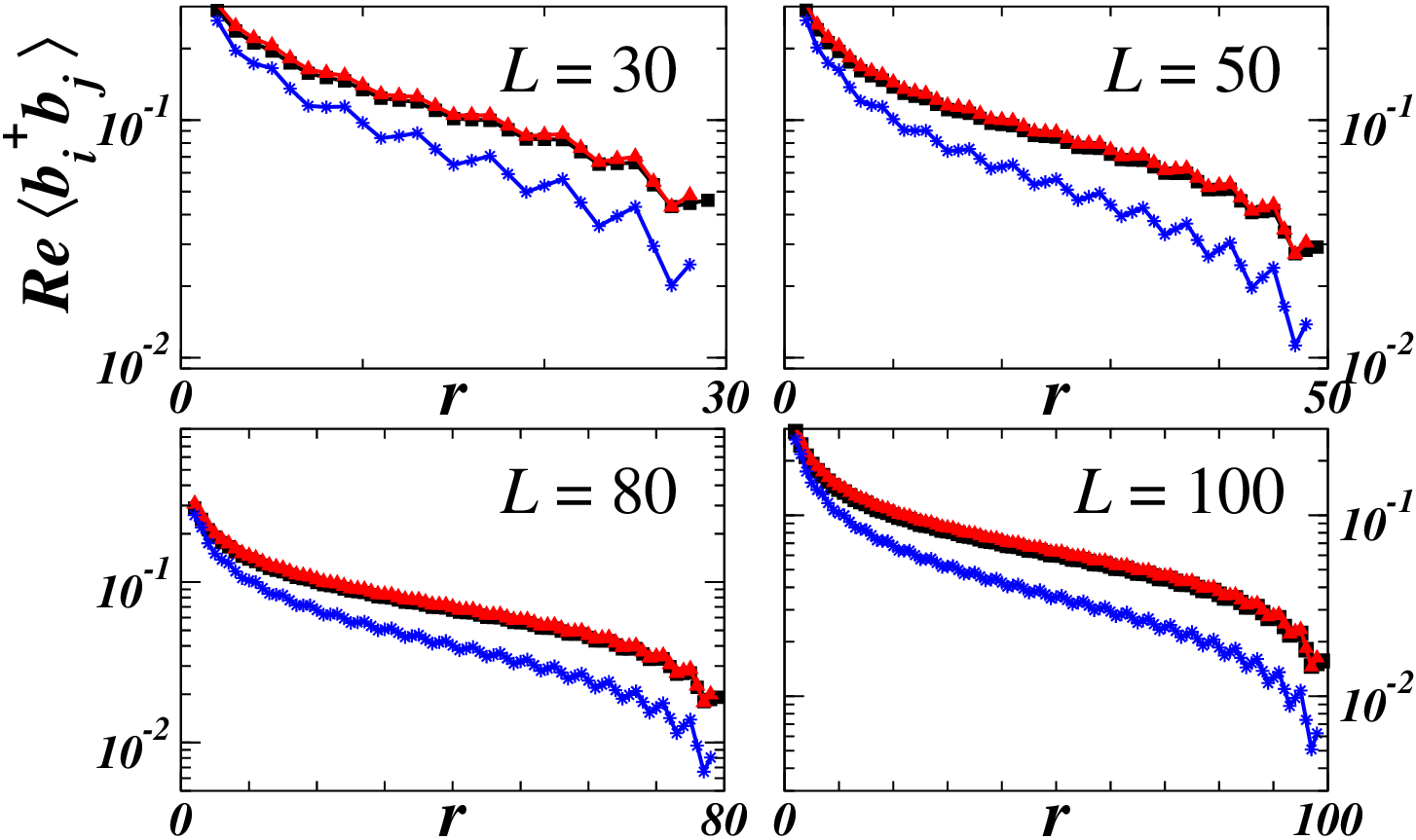}
  \caption{Real part of $\langle \hat{b}_i^\dag \hat{b}_j \rangle$ 
    {\it vs.} $r=\left| i-j \right|$ for the sizes and the parameters of 
    Fig.~\ref{FIGURE6}.}
  \label{FIGURE7}
\end{figure}

\begin{figure}[t]
  \includegraphics[width=\columnwidth]{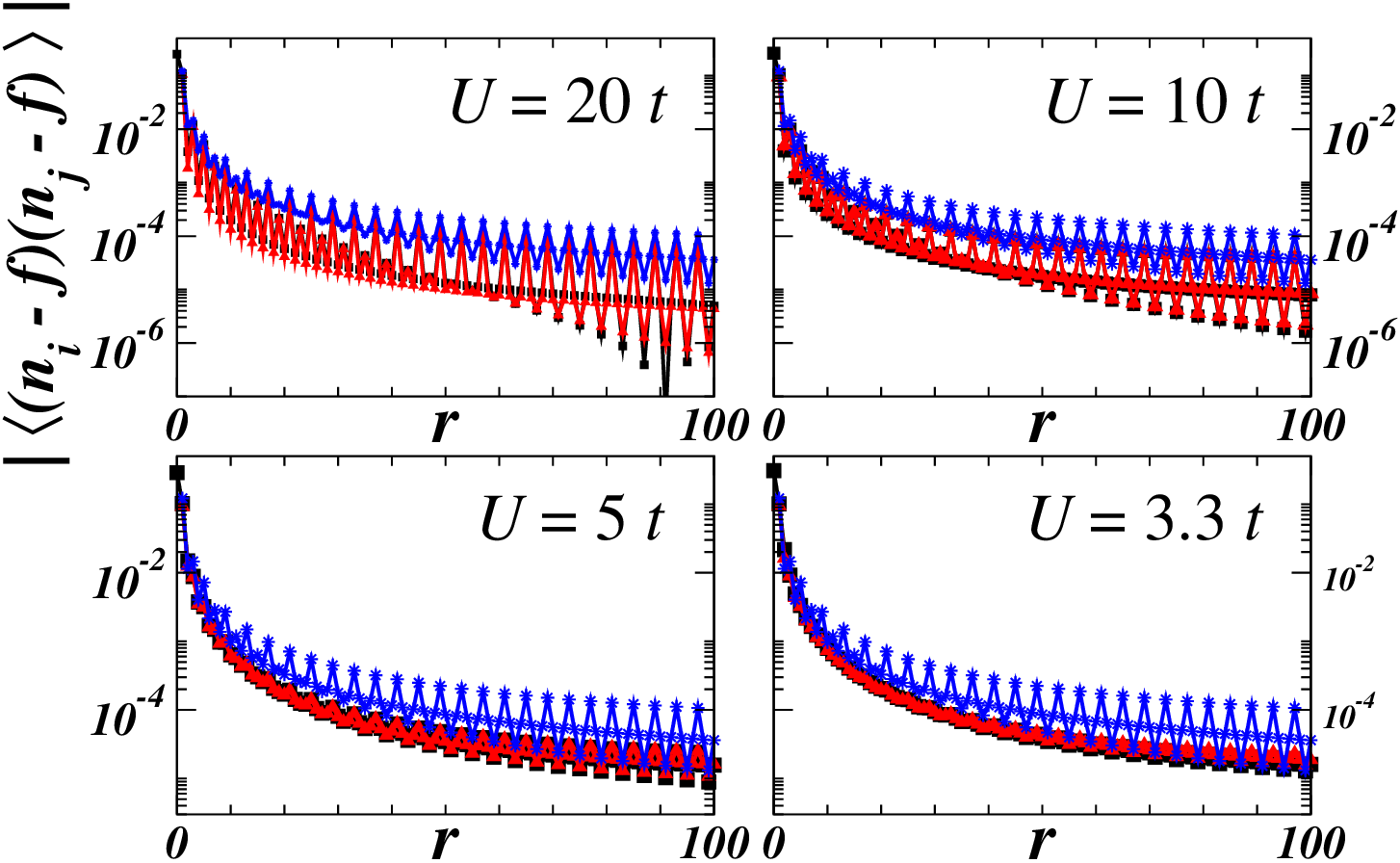}
  \caption{Density-density correlations 
    $\left| \langle \left(n_i-f \right)\left(n_j-f \right)\rangle \right|$ 
    {\it vs.} $r=\left| i-j \right|$ for different values 
    of $U/t = 20, 10, 5, 3.3$, corresponding, respectively, to  
    $J/U=0.1,0.2,0.4,0.6$ (with $V=0.5t$, $f=0.5$, $L=150$).}
\label{FIGURE8}
\end{figure}

\begin{figure}[t]
  \includegraphics[width=\columnwidth]{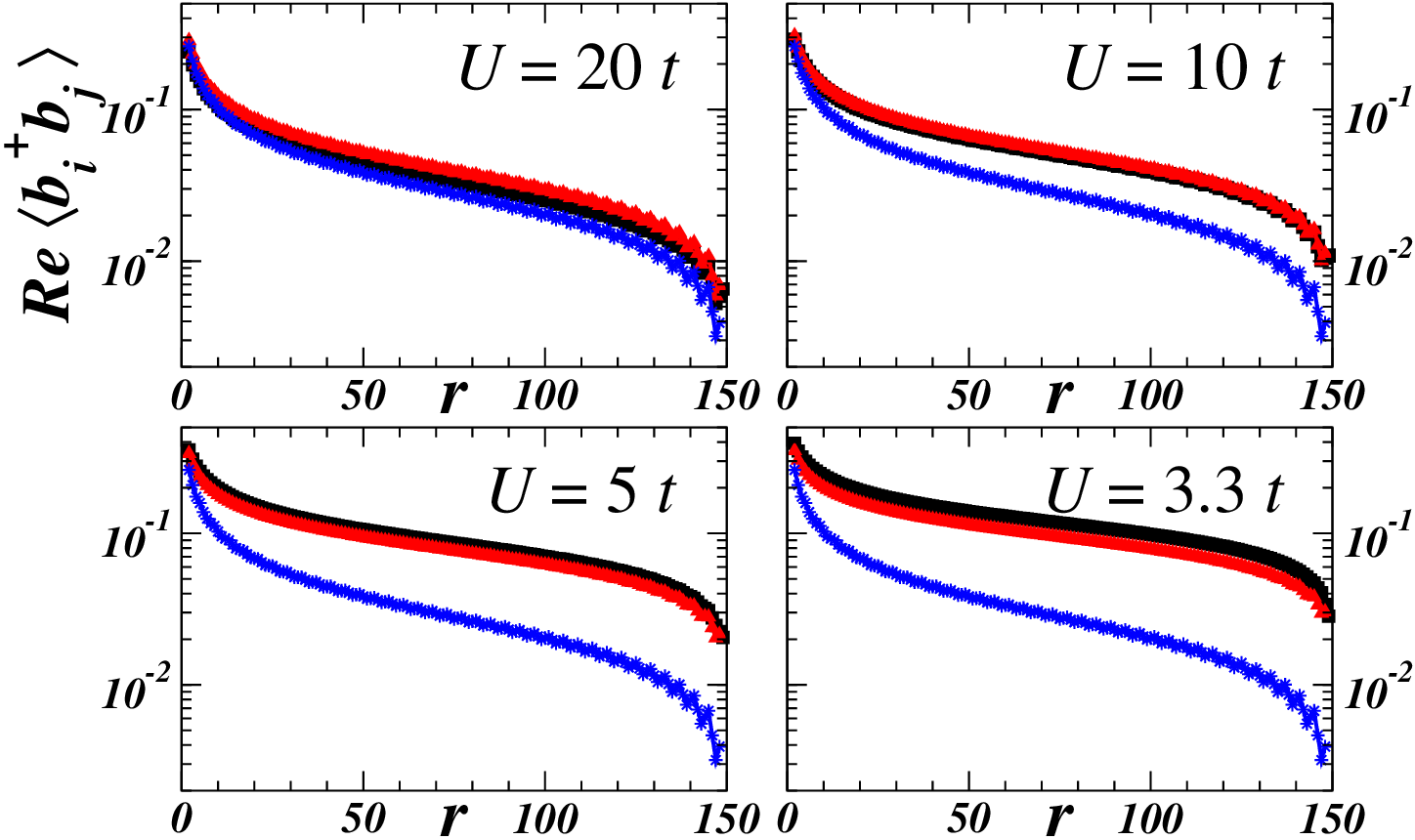}
  \caption{Real part of $\langle \hat{b}_i^\dag \hat{b}_j \rangle$ 
    {\it vs.} $r=\left| i-j \right|$ for the $U$ values and the parameters of Fig.~\ref{FIGURE8}.}
  \label{FIGURE9}
\end{figure}

The correlators $\langle \Phi_0 | \left(n_i-f \right)\left(n_j-f \right) | \Phi_0 \rangle$ 
and $\langle \Phi_0 | b_i^\dag b_j | \Phi_0 \rangle$ are evaluated 
from the corresponding XXZ quantities using respectively 
Eqs.~(\ref{corco.12}) and~(\ref{corco.17}). 
They are plotted in Figs.~\ref{FIGURE2}-\ref{FIGURE9} as a function of $r=\left| i-j \right|$, 
with $i$ and $j$ such that~\cite{hikihara98} $i=(L-r+1)/2$, $j=(L+r+1)/2$ for odd $r$, 
and $i=(L-r)/2$, $j=(L+r)/2$ for even $r$ (for instance, for $L=100$ sites, 
$r=1$ corresponds to $i=50, \, j=51$; $r=2$ corresponds to $i=49, \, j=51$; 
$r=3$ corresponds to $i=49, \, j=52$, and so on).
The meaning of the various symbols is summarized in the following table:
\begin{center}
  \begin{tabular}{ l | l }
    filled squares (black)  & numerical BH results \\ \hline
    filled diamonds (green) &  XXZ results - analytical $a$, $b$, $c$ \\ \hline
    filled triangles (red)  &  XXZ results - numerical $a$, $b$, $c$  \\ \hline
    stars (blue)  &  infinite-$U$ results \\ \hline
    open circles (magenta) &  non-rotated operators  \\
  \end{tabular}
\end{center}

In Fig.~\ref{FIGURE2} we plot our results for the density-density correlations 
$\langle \left( n_i - f \right) \left( n_j - f \right) \rangle$ for a typical set 
of values, i.e. for $U=10t$, $V=0.5t$, $f=0.5$, corresponding to $J/U=0.2$. 
Black squares (joint by a black line as a guide for eye) are the density-density 
correlations evaluated in the BH model, red triangles (line) are the correlation 
functions $\langle s_i^z s_j^z \rangle$ in the ground-state of the XXZ chain 
with effective anisotropy given by Eq.~(\ref{map1.17}) 
and the $a$, $b$, $c$ constants numerically determined from DMRG simulations 
of the XXZ chain, while the green diamonds (line) correspond to $a$, $b$, $c$ 
analytically determined from Eqs.~(\ref{xxz_5}) and~(\ref{xxz_6_a}, \ref{xxz_6_b}, \ref{xxz_6_c}). 
We found that, up to numerical accuracy $\lesssim 10^{-5}$, 
results obtained analytically for the XXZ effective model 
are in excellent agreement with results of the density-density 
BH model even at small distance. 
Blue stars (line) display the XXZ Hamiltonian results in the $U \to \infty$ limit, 
with anisotropy $\Delta=V/J$ -- in that case the relative error is noticeably larger.

In Fig.~\ref{FIGURE3} we plot the off-diagonal correlations $\langle b_i^\dag b_j \rangle$ 
for the same set of values of the BH parameters as in Fig.~\ref{FIGURE2}. 
Also here one sees that the results obtained from the GW effective Hamiltonian 
$H_{\rm XXZ}^{\rm eff}$ are in much better agreement than the ones 
obtained using $H_{\rm XXZ}$ with $\Delta=V/J$, this happens even though 
$J/U$ is as low as $0.2$. 

To quantify the agreement between BH and XXZ results, we consider the absolute value 
of the relative error done in evaluating a correlator ${\cal C}(r)$ as the ground-state average 
of the corresponding operators in the BH model [${\cal C}_{\rm BH}(r)$], and in the XXZ model
[${\cal C}_{\rm XXZ}(r)$]. More precisely, we define  
\begin{equation}
  \delta{\cal C}(r)=\left| \frac{{\cal C}_{\rm BH}(r)-{\cal C}_{\rm XXZ}(r)}{{\cal C}_{\rm BH}(r)} \right| 
  \label{rel_err}
\end{equation}
focusing on ${\cal C}_{zz}(r) \equiv \left| \langle \left( n_i - f \right) \left( n_j - f \right) \rangle \right|$ 
and ${\cal C}_{xy}(r) \equiv {\rm Re} \big[ \langle  b_i^\dag b_j \rangle \big]$. 
To summarize the information on the relative error, we compute the average value 
$\delta_{\rm av} {\cal C}$ and the standard deviation 
of the relative error~(\ref{rel_err}) for a distance $r=\left| i-j \right|$ 
between a minimum value $r_{\rm min} =1$ (2) for $zz$ ($xy$) correlations,
and a maximum value $r_{\rm max} \sim 3L/5$.

The relative errors for the $zz$ and $xy$ correlation functions are plotted 
in Figs.~\ref{FIGURE4}-\ref{FIGURE5}: the error made using the GW $H_{\rm XXZ}^{\rm eff}$ 
is of the order of few percents (in agreement with $(J/U)^2=0.04$). 
At variance, the relative error made by using the XXZ model in the infinite-$U$ limit 
without applying the GW procedure is much larger, although the value of $J/U$ is not so large. 
Indeed, the error $\delta_{\rm av}$ is $\sim 300\%$ for the $zz$ correlations and 
$\sim 40\%$ for the $xy$ correlations (to be compared with $\sim 6\%$ and $\sim 3\%$ 
obtained from $H_{\rm XXZ}^{\rm eff}$). 
We checked that these results do not depend on the particular choice of $r_{\rm max}$: 
of course, when $r_{\rm max}$ is closer to $L$, the error is larger 
(especially for the density-density correlations) due to boundary effects.
From the data of Figs.~\ref{FIGURE4}-\ref{FIGURE5}, one also sees that, 
at short distance, it is larger than that at intermediate distances
(with $r$ being few units it is $\lesssim 10\%$). 
As expected, it decreases at the center of the chain $r \sim L/2$, while, 
close to the end of the chain $r \sim L$, it increases. 
We also observe that finite-size effects are less visible for $xy$ correlations. 

The agreement between numerical and analytical results 
turns out to be stable also if one takes chains with smaller sizes, as it is apparent
from Figs.~\ref{FIGURE6}-\ref{FIGURE7}, where we plot the $zz$ and $xy$ correlation 
functions for different $L$. 
The corresponding errors are given in the following table:
\begin{center}
  \begin{tabular}{ l | l | l | l | l  }
    $L$  & $\delta_{\rm av}^{(U)} {\cal C}_{zz}$ & $\delta_{\rm av}^{(\infty)} {\cal C}_{zz}$   &  $\delta_{\rm av}^{(U)} {\cal C}_{xy}$  & $\delta_{\rm av}^{(\infty)} {\cal C}_{xy}$ \\ \hline
    $30$  &  $0.12 \pm 0.08$  &  $1.7 \pm 0.9$ &  $0.04 \pm 0.01$ &  $0.26\pm 0.07$  \\ \hline
    $50$  &  $0.10 \pm 0.06$  &  $2.0 \pm 1.0$ &  $0.04 \pm 0.01$ &  $0.29 \pm 0.08$ \\ \hline
    $80$  &  $0.09 \pm  0.05$ &  $2.4 \pm 1.0$ &  $0.04 \pm 0.01$ &  $0.33 \pm 0.08$ \\ \hline
    $100$ &  $0.08 \pm 0.05$  &  $2.5 \pm 1.1$ &  $0.04 \pm 0.01$ &  $0.35 \pm 0.08$ \\ \hline
    $150$ &  $0.07 \pm 0.04$  &  $2.9 \pm 1.2$ &  $0.04 \pm 0.01$ &  $0.38 \pm 0.08$ \\
  \end{tabular}
\end{center}
where for simplicity $\delta_{\rm av}^{(U)}$ ($\delta_{\rm av}^{(\infty)}$) denotes the average error 
for the XXZ correlators at finite-$U$ (infinite-$U$ limit) 
with (without) the GW procedure. We see that, for the density-density $zz$ correlations, 
the average error increases when the size $L$ decreases.

In Figs.~\ref{FIGURE3} and~\ref{FIGURE5} we also plotted (magenta circles and lines) 
the results obtained according to Eq.~(\ref{corco.1}), where we took 
${\cal O}_{\rm XXZ} ={\cal O}_{\rm BH}$ and not $ {\cal O}_{\rm XXZ} ={\bf S}^\dag {\cal O}_{\rm BH} {\bf S}$. 
Indeed, as stressed in Ref.~\onlinecite{kehrein06}, solving the equation for ${\bf S}$ amounts 
to perturbatively find a transformation enabling to block-diagonalize $H_{\rm BH}$. 
The ground-state of $H_{\rm BH}$ changes accordingly: if one wants to compute 
expectation values of certain operators in the BH model, one has to rotate the chosen 
operator according the ${\bf S}$ transformation - in other words, physical quantities 
in the effective theory are not simply the expectation values of the operators 
in the projected subspace: this guarantees the unitarity of the procedure.
An example is already provided in Ref.~\onlinecite{delannoy05} for the computation 
of the staggered magnetization in the 2D Fermi-Hubbard model with large-$U$ effective spin models. 
In Appendix~\ref{appop} we give details on the the explicit computation of the GW rotation for the 
operators $b_i^\dag b_j$ and $\left( n_i - f \right) \left( n_j - f \right)$. 
We remark that, while for density-density correlation functions 
($zz$ correlations in the XXZ model) magenta circles coincide with black squares, 
this is not the case for $\langle b_i^\dag b_j \rangle$ ($xy$ planar correlations in the XXZ model). 

In Figs.~\ref{FIGURE8}-\ref{FIGURE9} we plot the $zz$ and $xy$ correlation 
functions for different values of $U$: 
in these plots the ratio $J/U$ ranges from $0.1$ to $0.6$. 
As expected, one sees that for $J/U=0.1$ the relative error made by 
the infinite-$U$ results 
is not very large ($\sim 10\%$ for $\langle b_i^\dag b_j \rangle$ correlations), 
but, as soon as $J/U \gtrsim 0.2$, it is already well visible. 
The relative error made by using the effective $H_{\rm XXZ}^{\rm eff}$ 
turns out to be rather small even for $J/U=0.6$, where the relative error on 
density-density correlations is only $\approx 7\%$, 
while for $\langle b_i^\dag b_j \rangle$ correlations it is $\approx 15\%$~\cite{nota_4}.

\subsection{Antiferromagnet and domain ferromagnet in the 1D Bose-Hubbard model}   \label{transition}

The XXZ model is gapless and critical for $-1 \leq \Delta \leq 1$,  
antiferromagnetic for $\Delta > 1$ and ferromagnetic for $\Delta<-1$: 
in the latter ferromagnetic phase, all the spins are aligned. However, 
the BH at half-integer filling maps into the effective XXZ chain~(\ref{map1.18}) 
supplemented by the condition that the total spin is vanishing: 
therefore we expect that, in the BH model at $\Delta_{\rm eff}<-1$, domain walls form 
separating regions with ``up'' spins (i.e, with $f+1/2$ particles per site) 
and regions with ``down'' spins (i.e., with $f-1/2$ particles per site). 
At variance, at $\Delta_{\rm eff} >1$ the staggered magnetization 
becomes non vanishing: in the bosonic BH language the antiferromagnetic state 
corresponds to the ``charge-checkerboard ordered state'' $| f+1/2, f-1/2, f+1/2, f-1/2, \cdots \rangle$. 

This shows that, consistently with the XXZ representation of the BH model 
at half-integer filling, at finite $U$ one can realize the transition between 
the spin-liquid and the N\'eel-Ising antiferromagnetic phase of the XXZ model 
(superfluid to charge-density-wave phase transition of the BH model), as well as the
transition between the spin-liquid and the domain ferromagnetic Ising phase of the XXZ model 
(superfluid to domain Mott-insulating phase transition of the BH model)~\cite{nota_3}. 
Since the former transition sets in at $\Delta_{\rm eff} = 1$ and 
the latter one at $\Delta_{\rm eff} = -1$, using Eq.~(\ref{map1.17}) 
for $\Delta_{\rm eff}$ allows to determine the corresponding 
phase boundaries in terms of the parameters of the BH Hamiltonian. 

A complete discussion of the phase diagram of the BH chain in presence of  
nearest-neighbour interactions is provided in Ref.~\onlinecite{kuhner00}: 
here we just focus on the half-integer BH chain with parameters chosen so 
as to lie close to $\Delta_{\rm eff} = \pm 1$, in order to show that the effective XXZ representation 
given in this paper also provides a good description of these transitions. 

\begin{figure}[t]
  \includegraphics[width=\columnwidth]{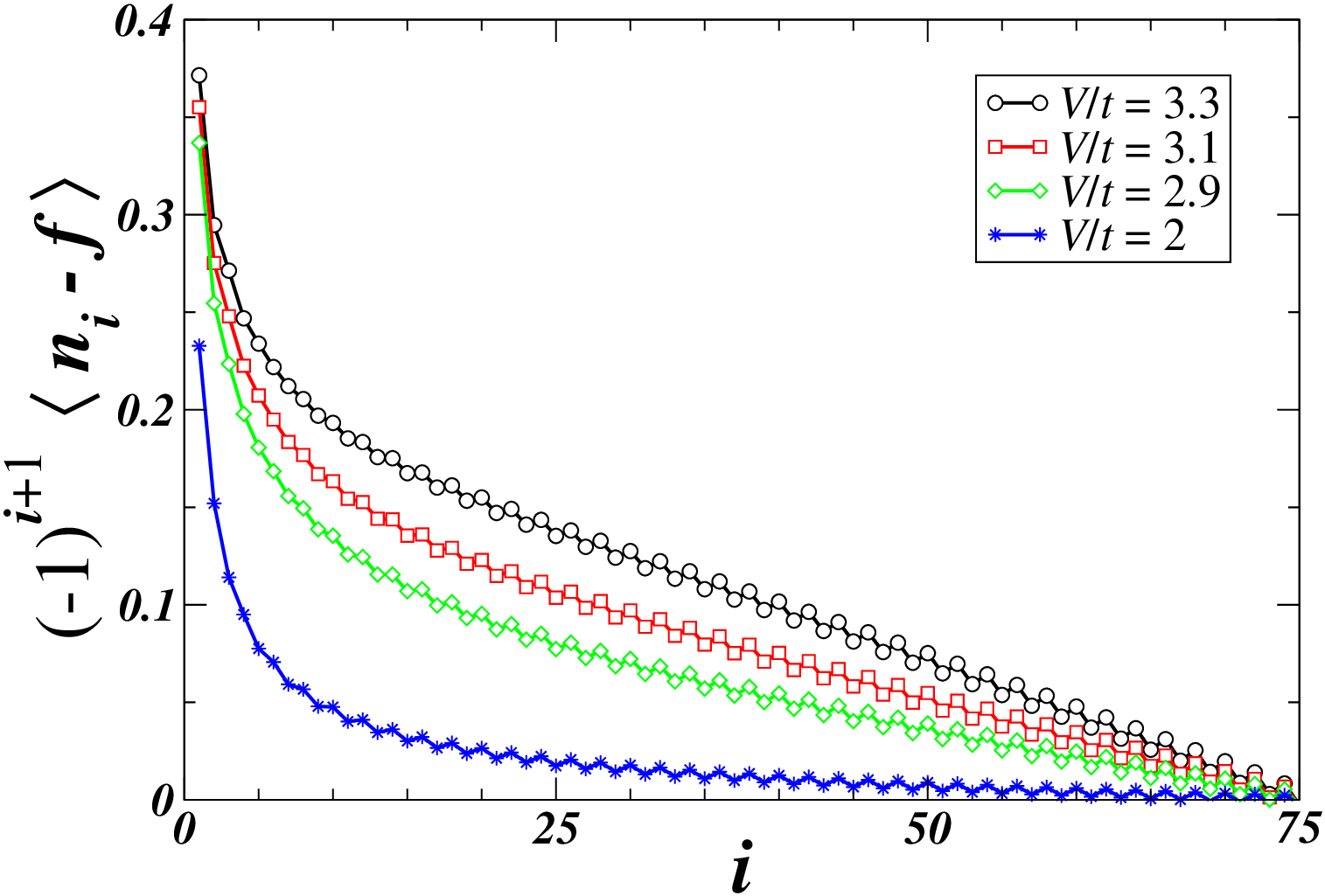}
  \caption{Plot of $(-1)^{i+1} \langle n_i-f \rangle$ 
    {\it vs.} $i$ numerically computed in the BH chain for $U=10t$, 
    $f=0.5$ and $L=150$ for different values of $V$: from top to bottom   
    $V/t=3.3$ (black circles), $V/t=3.1$ (red squares), 
    $V/t=2.9$ (green diamonds) and $V/t=2$ (blue stars) 
    corresponding, respectively, to $\Delta_{\rm eff}=1.09, 1.00, 0.91, 0.52$.}
  \label{FIGURE10}
\end{figure}

For the spin-liquid/ferromagnetic transition, we studied 
the BH chain with open boundary conditions varying $V$ (similar results 
are obtained varying $t$) and we plot in Fig.~\ref{FIGURE10} 
the expectation value of $(n_i-f)$ as a function of the position 
$i$ along the chain. We observe that, as a consequence of the open 
boundary conditions, a magnetic field proportional to $V$ 
on the two boundaries (i.e., at $i=1$ and $i=L$) appears, whose 
effect close to the boundaries is clearly visible in the figure. 
Computing the quantity ${\cal N}=\sum_r (-1)^{i-j} \langle (n_i-f) (n_j-f) \rangle$, 
one sees that it significantly increases around a critical value $\Delta_{\rm eff}^{AF}$. 
From the numerical data for the BH model shown in Fig.~\ref{FIGURE10} one may estimate 
$\Delta_{\rm eff}^{AF} \sim 1.05$, in good agreement 
with the analytical value $\Delta_{\rm eff}^{AF}=1$~\cite{nota_5}.
We notice that a better estimate of $\Delta_{\rm eff}^{AF}$ could be performed 
by adding a magnetic field in the boundaries to compensate 
the boundary magnetic fields arising from the open boundary conditions.
 
\begin{figure}[t]
  \includegraphics[width=\columnwidth]{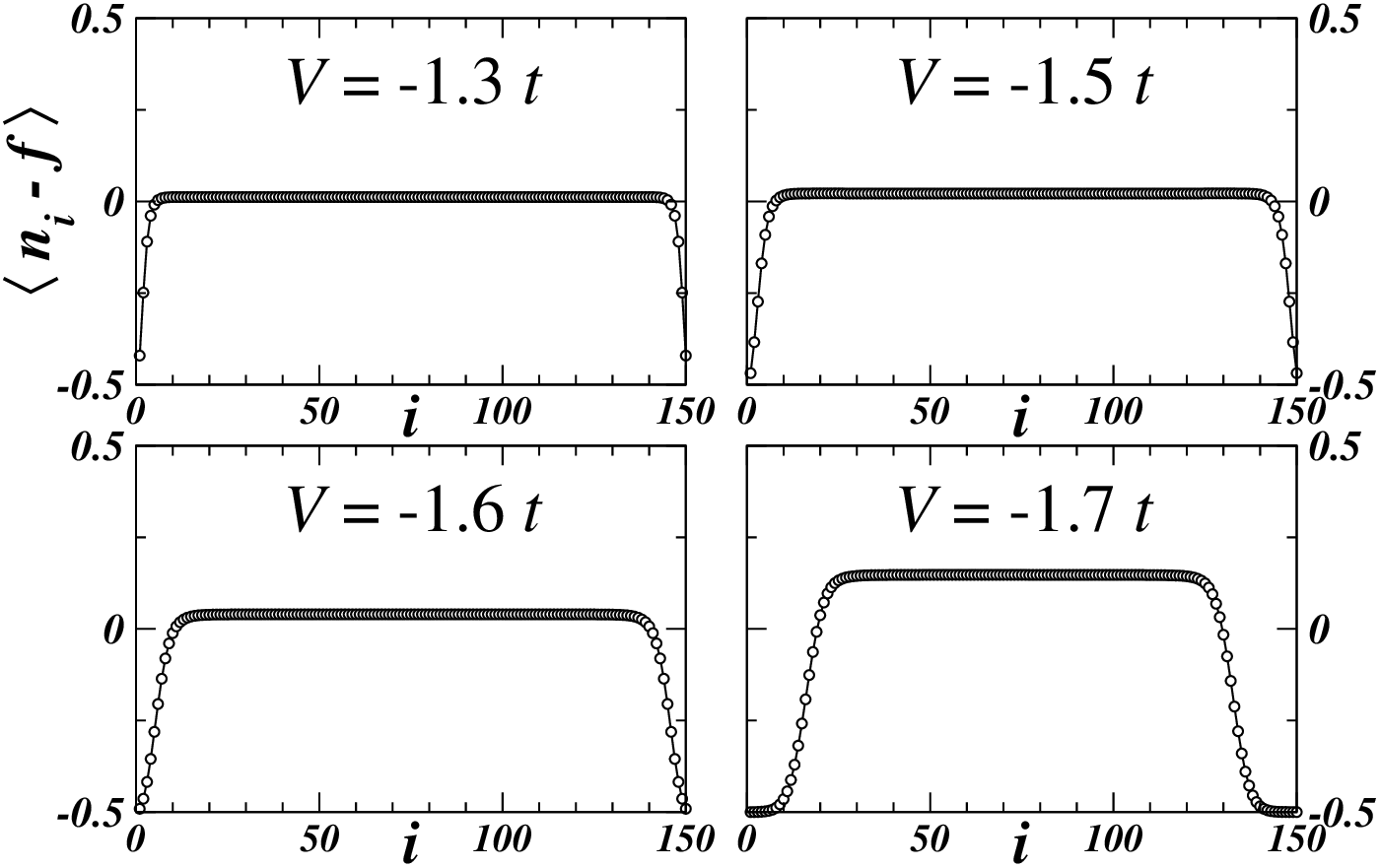}
  \caption{Plot of $\langle n_i-f \rangle$ 
    {\it vs.} $i$ numerically computed in the BH chain for $U=10t$, 
    $f=0.5$ and $L=150$ for different values of $V$: 
    $V/t=-1.3, -1.5, -1.6, -1.7$, corresponding, respectively, to
    $\Delta_{\rm eff}=-0.91, -1.00, -1.04, -1.08$.} 
  \label{FIGURE11}
\end{figure}

\begin{figure}[t]
  \includegraphics[width=\columnwidth]{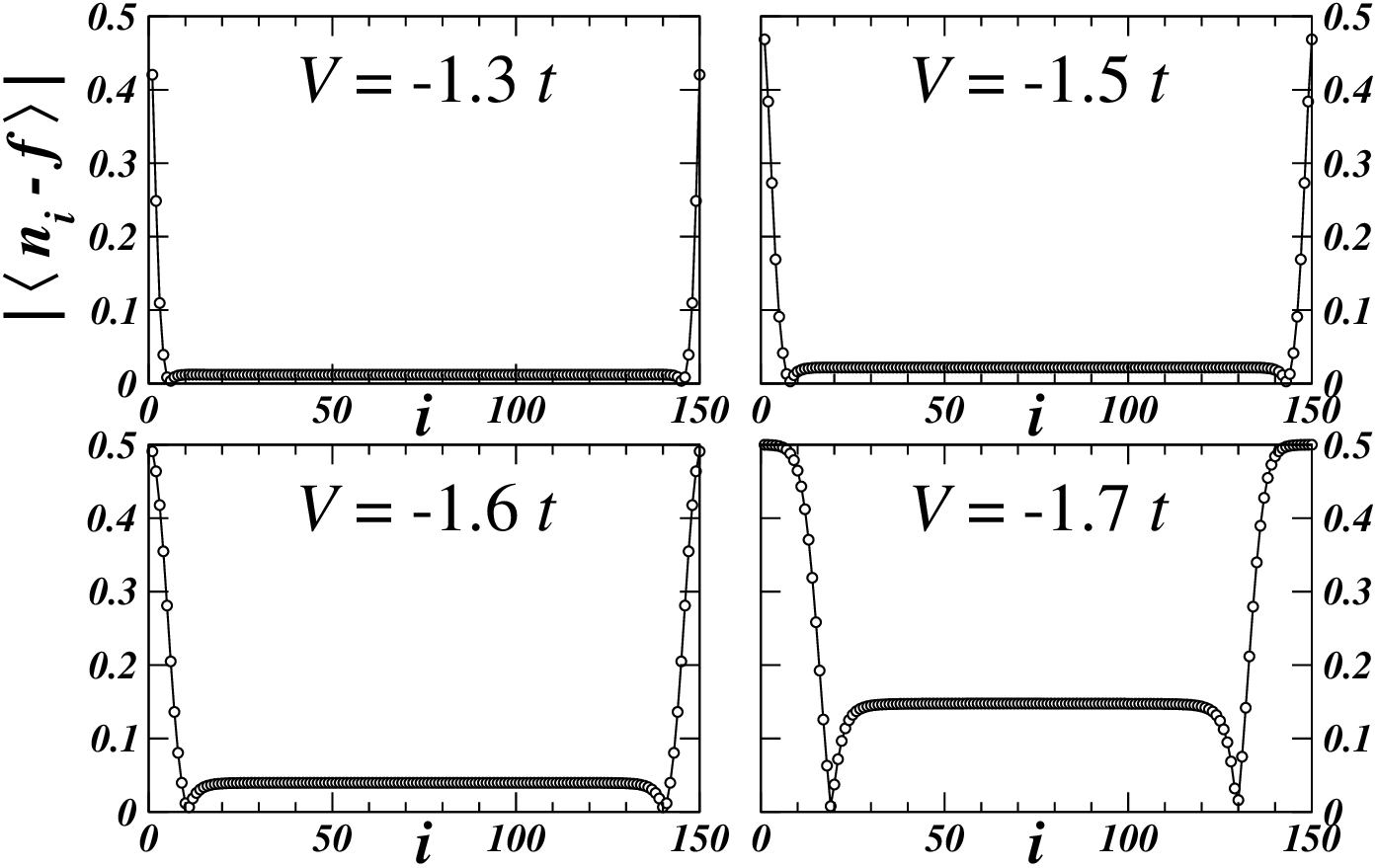}
  \caption{Plot of the modulus of $\langle n_i-f \rangle$ 
    {\it vs.} $i$ computed in the BH chain for the same parameters of 
    Fig.~\ref{FIGURE11}: $V/t=-1.3, -1.5, -1.6, -1.7$. The corresponding 
    averages are $\approx 0.02, 0.04, 0.06, 0.21$.}
  \label{FIGURE12}
\end{figure}

Regarding the domain ferromagnet/superfluid transition, we performed numerical simulations 
on the BH model with parameters chosen such that $\Delta_{\rm eff}$ is close to $-1$ 
(see Figs.~\ref{FIGURE11}-\ref{FIGURE12}). 
In Fig.~\ref{FIGURE11} we plot $\langle n_i-f \rangle$ as a function of the position $i$: 
one sees that the expectation value of the spin is constant and it changes sign close 
to the edges of the chain in order to satisfy the constraint on the 
number conservation. For this reason we then plot the modulus of the same quantity 
in Fig.~\ref{FIGURE12}: since the average of the $s_i^z$ expectation values 
is of course zero, to determine the transition point from 
BH numerical data we consider the averaged quantity $\sum_{i=1}^L |\langle n_i -f\rangle|$ 
(e.g., for the different values of $V$ shown in Fig.~\ref{FIGURE12}, such quantity is reported 
in the caption). From these data one can estimate that the domain ferromagnet is occurring at 
$\Delta_{\rm eff}^{F} \sim -1.02$, with an error of few percent with respect 
to the analytical result $\Delta_{\rm eff}^{F}=-1$~\cite{nota_5}. 
Notice that the error made by using $H_{\rm XXZ}$ with $\Delta=V/J$ 
in the infinite-$U$ limit is $\approx 20\%$: 
as expected, the errors made in using the infinite-$U$ results 
are generally smaller when one deals with global quantities.

\section{Concluding remarks}   \label{concl}

In this paper we studied an XXZ representation of the Bose-Hubbard chain 
at half-integer filling for finite on-site interaction energy 
$U$. The effective XXZ model is obtained in two steps: first, we used  
a similarity renormalization group procedure amounting 
to solve perturbatively up to the order $(t/U)^2$ 
the exact equation for the operator block-diagonalizing the Bose-Hubbard model. 
The resulting spin-1/2 effective Hamiltonian is then recast 
as a XXZ spin-1/2 Hamiltonian with pertinently redefined coupling 
and anisotropy parameters. 

We use this mapping to provide analytical estimates of the correlation 
functions of the Bose-Hubbard model at half-integer filling and finite 
$U$. We then compared these analytical results with the outcomes 
of the numerical DMRG evaluation of the Bose-Hubbard correlation functions. 
We found that the agreement is very good, 
also for $J/U$ rather large ($\sim 0.5$) and for small number of sizes 
($L \sim 30$). Such a good agreement is not achieved,  
even for $J/U$ relatively small ($\sim 0.1$), if one uses 
the XXZ Hamiltonian $H_{\rm XXZ}^{(0)}$ with $J=2t(f+1/2)$ and 
$\Delta=V/J$ corresponding to the infinite-coupling limit of the Bose-Hubbard model. 
The transitions predicted at $\Delta_{\rm eff}=\pm 1$ for the XXZ chain are 
as well compared with Bose-Hubbard results, and a good agreement is found.

Since the BH model at half-integer filing is not integrable 
or exactly solvable, it is quite valuable to have analytical estimates 
for its correlation functions. Besides its mathematical interest, 
we stress out that our results can be viewed from a two-fold point of view: 
on one side, we use known results from the (integrable) XXZ model 
to construct with high accuracy correlation functions of the 
Bose-Hubbard model. On the other side, the Bose-Hubbard chain at half-filling 
and at finite $U$ may be seen as a quantum simulator of the XXZ chain.
Finally, our results could be relevant towards extending to
the BH model the analysis of nonequilibrium steady state in the XXZ chain
performed in Ref.~\onlinecite{prosen}.

In our approach, the effect of an harmonic trap results in a 
locally varying magnetic field: we feel that it would be interesting 
to compare the results stemming from an XXZ-based approach with the ones 
known in literature for hard- and soft-core bosons in harmonic traps 
in the scaling limit~\cite{campostrini09}. In this paper we focused on the half-integer 
filling Bose-Hubbard model, but deviations from such filling could be 
easily accounted with the introduction of a magnetic field. 
We stress that the similarity Hamiltonian renormalization procedure 
could also be applied to bosonic ladders~\cite{carrasquilla10} and at 
integer filling, where a spin-1 model is found in the infinite-$U$ limit. 

The large-$V$ effects of edge magnetic field could also be studied, 
following the results known for the XXZ chain~\cite{affleck98}: 
we observe that, for open boundary conditions and finite $V$, two boundary 
magnetic field terms $-B_b (s_1^z + s_L^{z})$, with $B_b \propto V$, 
emerge in the XXZ effective Hamiltonian~\cite{nota_6}. 
Since a magnetic field at the edge induces corrections to the average value of $s_i^z$ 
decreasing as a power law~\cite{affleck98}, these corrections 
are not only expected, but could be also worth the effort of future investigation.  

\acknowledgments

We would like to thank F. Becca, L. Campos-Venuti, 
F. Essler, A. Ferraz, V. Korepin, F. Minardi, M. M\"uller, 
G. Santoro and A. Smerzi for very useful discussions. 
A.T. acknowledges kind hospitality from IIP-UFRN (Natal), where 
part of this work was performed.
D.R. acknowledges financial support from EU through the project SOLID.

\appendix

\section{Perturbative solution of the GW equation}   \label{appgw}

In this Appendix we show how use Eq.~(\ref{simi_7}) to determine ${\bf a}$ to {\it first order} 
in $H_I$, that is ${\bf a_1}$.
To this order, one gets
\begin{equation}
  {\cal P} \left\{ H_I + [ H_0 , {\bf a_1} ] \right\}  \left( {\bf I} - {\cal P} \right) = 0,
  \label{simi_9}
\end{equation}
which may be solved by setting
\begin{eqnarray}
&& {\cal P}  {\bf a_1}  \left( {\bf I} - {\cal P} \right)  = \nonumber \\
 && {\cal P}  H_I  \left( {\bf I} - {\cal P} \right)  \big\{ - {\cal P} 
  H_0 {\cal P} + \left( {\bf I} - {\cal P} \right) H_0 
  \left( {\bf I} - {\cal P} \right)  \big\}^{-1} \nonumber \\
  & & +  [ {\cal P} H_0 {\cal P}  , {\cal P}  {\bf a_1}  \left( {\bf I} - {\cal P} \right) ]
\nonumber \\
&&\times    \big\{ - {\cal P} H_0 {\cal P}  + \left( {\bf I} - {\cal P} \right) H_0 
  \left( {\bf I} - {\cal P} \right) \big\}^{-1}.
  \label{simi_10}
\end{eqnarray}
Up to term that are second order in $ t \bar{n} / U$, 
we may make the approximation  $ {\cal P}  H_0 {\cal P}  \approx {\cal E}_0 [ \bar{n} ] {\bf I}$, 
with  ${\cal E}_0 [ \bar{n} ] = L \left\{ \frac{U}{2} \bar{n} ( \bar{n} - 1 ) + V \bar{n}^2 \right\}$,
which implies $[ {\cal P} H_0 {\cal P} , {\cal P} {\bf a}  \left( {\bf I} - {\cal P} \right) ] = 0$. 
As a result, we get
\begin{eqnarray}
&&  {\cal P}  {\bf a_1}  \left( {\bf I} - {\cal P} \right) =  {\cal P} H_I  \left( {\bf I} - {\cal P} \right) \nonumber \\
&& \times \left\{ - {\cal P} H_0 {\cal P}  + \left( {\bf I} - 
  {\cal P} \right) H_0 \left( {\bf I} - {\cal P} \right)  \right\}^{-1}.
  \label{simi_11}
\end{eqnarray}
Using the fact that ${\bf a}$ is antihermitean, from Eq.~(\ref{simi_11}) one obtains
\begin{eqnarray}
&&  {\bf a_1} = {\cal P}  H_I  \left( {\bf I} - {\cal P} \right) \nonumber \\ && \times 
  \big\{ - {\cal P} H_0 {\cal P}  + \left( {\bf I} - {\cal P} \right) H_0 \left( {\bf I} - 
  {\cal P} \right) \big\}^{-1} -  \nonumber \\
&& \big\{ - {\cal P} 
  H_0 {\cal P}  + \left( {\bf I} - {\cal P} \right) H_0 \left( {\bf I} 
  - {\cal P}  \right) \big\}^{-1} \nonumber \\
&& \times  \left( {\bf I} - {\cal P} \right) H_I {\cal P}.
  \label{simi_12}
\end{eqnarray}

\section{GW transformation of operators}   \label{appop}

An advantage of the GW procedure is that it may be easily applied to single-boson operators: 
in particular, we are interested in the average values of 
the operators ${\cal M}_{i,j}^\perp$ and ${\cal M}_{i,j}^z $ defined in Eqs.~(\ref{corko.a}) and (\ref{corko.b}).
Since ${\bf a_1}$ is fully off-diagonal and ${\cal P}  {\cal M}_{i,j}^z \left( {\bf I} - {\cal P}  \right)  = 0$, 
if one approximates ${\bf T} $ with ${\bf a_1}$, one obtains ${\bf S}^\dag {\cal M}_{i,j}^z {\bf S} = {\cal M}_{i,j}^z $.
Instead, acting onto  ${\cal M}_{i,j}^\perp $ gives raise to a more complicated expression: 
expressing the final result in terms of spin-1/2 variables, one obtains
\begin{widetext}
  \begin{eqnarray}
    {\cal P}  {\bf S}^\dag {\cal M}_{i,j}^\perp  {\bf S} {\cal P} & \approx &
    \delta_{ | i - j | , 1} \: \frac{t \bar{n} ( \bar{n} + 2 )}{U} \left( \frac{1}{2} - s_{i + 1}^z 
    \right) \left( \frac{1}{2} + s_{i }^z  \right) \\
    & + & \frac{t ( \bar{n} + 2)(\bar{n} + 1 ) }{U}  
    \biggl\{ s_{i + 1}^- s_{j}^+ \left( \frac{1}{2} + s_i^z \right) 
    + s_{i - 1}^- s_{j}^+ \left( \frac{1}{2} + s_i^z \right) 
    +  s_{i }^- s_{j+1}^+ \left( \frac{1}{2} + s_{j}^z \right) 
    +  s_{i}^- s_{j -1}^+ \left( \frac{1}{2} + s_{j}^z \right) \biggr\}
    \nonumber \\
    & + & \frac{t ( \bar{n} + 2)(\bar{n} + 1 ) }{2 U} \biggl\{ s_{i }^- s_{j -1}^+ \left( \frac{1}{2} - s_i^z \right) 
    + s_{i }^- s_{j +1}^+ \left( \frac{1}{2} - s_i^z \right)+ s_{i -1}^- s_{j}^+ \left( \frac{1}{2} - s_{j}^z \right) 
    +  s_{i  +1}^- s_{j}^+ \left( \frac{1}{2} - s_{j}^z \right)  \biggr\}. \nonumber
    \label{corco.16}
  \end{eqnarray}
\end{widetext}

We observe that due to the constraint on the fixed total particle number $N$, 
the total magnetization in any of the physical states
of $H_{\rm eff} = H_{\rm XXZ}^{(0)} + H_{\rm diag}^{(1)} + H_{\rm offd}^{(1)}$ is zero: 
since $H_{\rm eff}$ contains no terms breaking the parity symmetry 
($s_i^\alpha \to - s_i^\alpha$), its ground-state $| \Psi_0 \rangle $ is
 nondegenerate and, thus, it must be parity invariant. 
As a consequence, the average of any product of three spin-1/2 operators must necessarily give $0$, 
greatly simplifying the calculation of the ground-state average of the operator.

Using this result one can obtain a simplified expression for 
$\langle \Phi_0 | {\cal M}_{i,j}^\perp  | \Phi_0 \rangle$ at 
${\cal O} \left( \frac{t^2 \bar{n}^2}{U^2} \right)$: 
\begin{widetext}
\begin{eqnarray}
  \label{corco.16bis}
  && \langle \Phi_0 | {\cal M}_{i,j}^\perp  | \Phi_0 \rangle  \approx 
  \delta_{ | i - j | , 1} \: \langle \Psi_0 |  \left( \frac{1}{2} - s_{i + 1}^z 
  \right) \left( \frac{1}{2} + s_{i }^z  \right)| \Psi_0 \rangle \\
  & & +
  \frac{t ( \bar{n} + 2)(\bar{n} + 1 ) }{U} 
 \langle \Psi_0 |  
  \biggl\{ s_{i + 1}^- s_{j}^+ \left( \frac{1}{2} + s_i^z \right) 
  + s_{i - 1}^- s_{j}^+ \left( \frac{1}{2} + s_i^z \right)  
  +  s_{i }^- s_{j+1}^+ \left( \frac{1}{2} + s_{j}^z \right) 
  +  s_{i}^- s_{j -1}^+ \left( \frac{1}{2} + s_{j}^z \right) \biggr\}
  {\cal P} | \Psi_0 \rangle 
  \nonumber\\  & &
  + \frac{t  \bar{n} (\bar{n} + 1 ) }{U}  \langle \Psi_0 |  {\cal P} 
  \biggl\{ s_{i }^- s_{j -1}^+ \left( \frac{1}{2} - s_i^z \right)
  + s_{i }^- s_{j +1}^+ \left( \frac{1}{2} - s_i^z \right) + 
  + s_{i -1}^- s_{j}^+ \left( \frac{1}{2} - s_{j}^z \right)  
  +  s_{i  +1}^- s_{j}^+ \left( \frac{1}{2} - s_{j}^z \right)  \biggr\}
  {\cal P} | \Psi_0 \rangle \nonumber
\end{eqnarray}
\end{widetext}
Since any product of three spin-1/2 operators must necessarily give $0$, 
then Eq.~(\ref{corco.16bis}) gives Eq.~(\ref{corco.17}) reported in the main text.


\begin{thebibliography}{XX}
\bibitem{nota_0} 
  Since only eigenstates with zero total spin along the $z$ axis are physically meaningful 
  (due to the constraint on the total number of particles), in the following we use
  the formulas of Ref.~\onlinecite{hikihara98} at vanishing applied magnetic field.
\bibitem{nota_2}
  Although the contribution to the total energy arising from magnetic fields vanishes 
  on the physical states due to the constraint on the particle number, 
  in the intermediate calculations one has to retain it when computing 
  the redefined XXZ parameters.
\bibitem{nota_3} Notice that, as discussed in the following, the constraint on the total number 
(implying that only states of the XXZ model with the z
component of the total spin equal to zero may be realized within
the BH model in the canonical ensemble) does not affect the spin 
liquid /  N\'eel-Ising phase transition since it just implies the existence of 
(at least two) ferromagnetic domain walls, in the ferromagnetic- Ising phase of the XXZ model.
\bibitem{nota_4}
  This result does not depend on the particular choice of $r_{\rm max}$.
\bibitem{nota_5}
  In passing we point out that, to obtain more accurate results on the phase transition points,
  a finite-size scaling of the data with $L$ is needed.
  Namely, one should estimate the value of $\Delta_{\rm eff}^{AF/F}$ for different sizes $L$,
  and then perform a fit of that quantity with $L$, such to extrapolate the thermodynamic limit.
  This analysis lies beyond our present purposes, which are to provide estimates of the transition 
  within a precision of $\approx 5 \%$.
\bibitem{nota_6} Notice that computing the renormalization of the effective XXZ
parameters via the Luttinger treatment done in Sec.~\ref{equivalence} with periodic
boundary conditions is enough to give excellent agreement with BH
numerical data, provided that, of course, the correlation functions
of the XXZ model are computed for open boundary conditions according
to Eqs.~(\ref{xxz_2}, \ref{xxz_4}).
\end{thebibliography}
\end{document}